\documentclass{emulateapj}
\usepackage{appendix}

\received{}
\accepted{}
\slugcomment{Accepted by ApJ, 23 January 2013}
\shorttitle{HI in Massive Gas-rich Mergers}
\shortauthors{Teng, Veilleux, \& Baker}

\begin{document}

\title{GBT Detection of Polarization-dependent HI Absorption and HI Outflows in Local ULIRGs and Quasars}

\author{Stacy H. Teng~\altaffilmark{1, 7}, Sylvain Veilleux~\altaffilmark{2, 3, 4, 5, 7}, \& Andrew J. Baker~\altaffilmark{6}}

\altaffiltext{1}{Observational Cosmology Laboratory, NASA Goddard Space Flight Center, Greenbelt, MD 20771, USA; stacy.h.teng@nasa.gov}
\altaffiltext{2}{Department of Astronomy, University of Maryland, College Park, MD 20742, USA}
\altaffiltext{3}{Joint Space-Science Institute, University of Maryland, College Park, MD 20742, USA}
\altaffiltext{4}{Astroparticle Physics Laboratory, NASA Goddard Space Flight Center, Greenbelt, MD 20771, USA}
\altaffiltext{5}{Max-planck-Institut f\"{u}r extraterrestrische Physik, Postfach 1312, D-85741 Garching, Germany}
\altaffiltext{6}{Department of Physics and Astronomy, Rutgers, the State University of New Jersey, Piscataway, NJ 08854, USA}
\altaffiltext{7}{NASA Postdoctoral Program Fellow}

\begin{abstract}

We present the results of a 21-cm HI survey of 27 local massive gas-rich late-stage mergers and merger remnants with the Robert C. Byrd Green Bank Telescope (GBT). These remnants were selected from the Quasar/ULIRG Evolution Study (QUEST) sample of ultraluminous infrared galaxies (ULIRGs; L$_{8-1000~\mu m} > 10^{12}$ L$_\odot$) and quasars; our targets are all bolometrically dominated by active galactic nuclei (AGN) and sample the later phases of the proposed ULIRG-to-quasar evolutionary sequence. We find the prevalence of HI absorption (emission) to be 100\% (29\%) in ULIRGs with HI detections, 100\% (88\%) in FIR-strong quasars, and 63\% (100\%) in FIR-weak quasars.  The absorption features are associated with powerful neutral outflows that change from being mainly driven by star formation in ULIRGs to being driven by the AGN in the quasars.  These outflows have velocities that exceed 1500~km~s$^{-1}$ in some cases.  Unexpectedly, we find polarization-dependent HI absorption in 57\% of our spectra (88\% and 63\% of the FIR-strong and FIR-weak quasars, respectively).  We attribute this result to absorption of polarized continuum emission from these sources by foreground HI clouds.  About 60\% of the quasars displaying polarized spectra are radio-loud, far higher than the $\sim$10\% observed in the general AGN population.  This discrepancy suggests that radio jets play an important role in shaping the environments in these galaxies.  These systems may represent a transition phase in the evolution of gas-rich mergers into ``mature'' radio galaxies.

\end{abstract}


\section{Introduction}
\label{sec:intro}


Recent theoretical work suggests that feedback from AGN plays an important role in establishing the present-day appearance of massive galaxies.  Major galaxy mergers at high redshifts may trigger powerful starbursts, lead to the formation of elliptical galaxies, and account for the growth of supermassive black holes \citep[e.g.,][]{sanders88, dimatteo05, springel05, hopkins05, hopkins08}.  In this merger-driven evolutionary scenario, two gas-rich progenitors collide to form a completely obscured ULIRG where substantial inflows of gas trigger starbursts and feed the central black hole(s).  As the system evolves, the obscuring gas and dust begin to disperse, bringing about the formation of a dusty IR-excess quasar and eventually a ``naked'' optical quasar.  Powerful winds, driven by either the starburst or the central black hole, seem to be required to quench star formation and to stop the growth of the black hole in order to explain the observed $M_{\rm BH} - \sigma$ relation \citep[e.g.,][]{murray05}.  

There is growing support for the existence of this type of ``negative feedback'' in luminous late-stage mergers.  Most galaxies with star formation rate (SFR) densities $\gtrsim$0.1~$M_\odot$~yr$^{-1}$~kpc$^{-2}$ show signs of outflow \citep[e.g.,][]{heckman02, vei05}.  These starburst-driven winds are often seen in ULIRGs, with mass outflow rates of $\sim$0.1--5 times the SFR and velocities $\sim$100--400~km~s$^{-1}$ \citep{rupke02, rupke05a, rupke05b, rupke05c, martin05, martin06}.  In late-stage mergers and pure Seyfert galaxies, however, these winds are sometimes driven by the AGN, reaching outflow velocities $\gtrsim$1000~km~s$^{-1}$ \citep{rupke05c, krug10}.  Similar AGN-driven outflows are also found in recently (re)started radio galaxies \citep[e.g.,][]{oosterloo00, morganti05b}.  Several studies have also shown that these outflows are often multi-phased, with powerful molecular, neutral, and ionized outflows having all been detected in the same object \citep[e.g.,][]{morganti07, fischer10, feruglio10, rupke11, sturm11}.  


In the present paper, we focus on the HI properties of the QUEST ULIRGs and quasars, 
searching for the presence and the driving source (AGN or starburst) of neutral outflows in merging systems as a function of merger stage.  
This paper is organized as follows.  In Section~\ref{sec:sample}, we discuss how our HI sample is selected.  In Section~\ref{sec:obs}, we describe our observational setup, data reduction, and spectral analysis.  We then examine the HI profiles of our targets in Section~\ref{sec:profiles} and the detection of polarized absorption (including a more
sophisticated spectral decomposition that boosts our number of absorption detections) in Section~\ref{sec:polar}.  Finally, in Section~\ref{sec:origin}, we discuss the possible origins of the HI outflows.  We conclude in Section~\ref{sec:sum} with a summary of our findings.  Throughout this paper, we adopt $H_0$ = 71 km~s$^{-1}$~Mpc$^{-1}$, $\Omega_M$ = 0.27, and $\Omega_\Lambda$ = 0.73 \citep{wmap}.

\section{Sample Selection} 
\label{sec:sample}

In order to assess the physical conditions and the dynamical evolution of the neutral gas in late-stage merger systems, we selected targets from the QUEST sample of ULIRGs and quasars that are bolometrically dominated by AGN.   The details of the QUEST sample selection are outlined in previous QUEST papers \citep[e.g.,][]{schweitzer06, netzer07, vei09a}.  Briefly, the original QUEST sample of ULIRGs begins with the 1-Jy sample, a flux-limited sample of 118 ULIRGs selected from a redshift survey of the {\it IRAS} faint source catalog \citep{kim98}.  To these, five well-studied local ULIRGs from the Revised Bright Galaxy Survey \citep[RBGS;][]{rbgs} were added given the availability of archival data from multiple observatories \citep[see][for details]{vei09a, vei12}.  The original quasar sample contains 25 quasars, including 24 Palomar-Green (PG) quasars from the Bright Quasar Sample \citep{pgqso} and B2~2201+31A which satisfies the $B$ magnitude criterion of \citet{pgqso}.  An additional nine PG quasars were added to the sample from archival data.  The selected quasars fall at the low end of the quasar $B$-band luminosity function.

The ULIRGs and quasars in the QUEST sample are well-matched in redshift ($z < 0.3$) and in near-IR imaging show tidal features that indicate merger origins \citep{vei06, vei09b}.  The entire QUEST sample spans the merger evolution sequence from binary starburst-dominated ULIRGs, to completely coalesced AGN-dominated ULIRGs, to dust-enshrouded quasars, to optical quasars whose dust has been cleared. 

Our HI sample is a subset of the QUEST targets.  To ensure quality spectra, we require $z < 0.12$; our 27 targets evenly sample coalesced ULIRGs\footnote{This category includes Mrk~273, the only source in our sample to have binary nuclei (see Appendix).} (9 targets), far-infrared(FIR)-strong quasars ($f_{\rm 60 \mu m}/f_{\rm 15 \mu m} > 1$; 10 targets), and FIR-weak quasars (8 targets).  The two classes of quasars have similar spectral energy distributions representative of AGN, with the FIR-strong quasars having an additional contribution to their FIR emission from intense star formation \citep{netzer07}.  The morphologies of the quasar hosts are in agreement with the picture that FIR-weak quasars are at a later evolutionary stage than FIR-strong quasars, following the (near) cessation of star formation \citep{vei09b}.  Thus, our targets sample the later stages of merger evolution from coalescence to the aftermath of significant star formation.  Table~\ref{tab:sample} summarizes the properties of our HI sample.

\section{Observations and Data}
\label{sec:obs}

\subsection{Observational Setup}

The targets were observed with the GBT Spectrometer in the L band between 2011 August and 2012 March during the $\sim$160 hours of time allocated for GBT programs 11B075 and 12A075 (PI: Teng).  The data were collected in 10 minute on-off source pairs (5 minutes per position) with 12.5 MHz bandwidth, two spectral windows centered at the same frequency (1420.4058 MHz), two linear polarization modes (XX and YY), and nine sampling levels.  To minimize the effects of radio frequency interference (RFI) on the data, we used the notch filter for sources with heliocentric recessional velocities below 17000~km~s$^{-1}$.  The total on-source time for each galaxy was determined during the observing runs based on the strength of the spectral features relative to the noise or up to a maximum of three hours.  Most sources were observed over multiple observing sessions.  The circumpolar starburst galaxy M82 was observed at the beginning of each observing session as a check for instrumental variations.

We also measured the continuum flux densities of our sample at 1.42~GHz using the AutoPeak procedure in order to check for time variability.  However, due to the plethora of continuum sources within the large GBT beam, the 5-$\sigma$ confusion limit is $\sim$100~mJy at this frequency.  Therefore, we were only able to obtain confident continuum measurements for five targets.  These values are in agreement with older values obtained with the Very Large Array (VLA) for their nuclear core flux densities \citep{condon02}, despite the much larger beam of the GBT.  For the remainder of sample, we use published values of the core continuum flux densities as measured with the VLA \citep[e.g.,][]{barvainis, nvss, rafter}.  

\subsection{Data Reduction}
\label{sec:redux}

The spectra were reduced using GBTIDL \citep{gbtidl}.  Individual five-second records showing large broadband harmonic RFI were flagged, and persistent RFI spikes that appear in the data were replaced by interpolation across the affected channels for both spectral windows and polarizations.  For each polarization, the data from the two spectral windows were averaged together, and a polynomial of order $\leq$5 was fitted over a range of $\sim$1000~km~s$^{-1}$ on each side of a galaxy's systemic redshift in the Hanning-smoothed and then decimated\footnote{The total number of channels in the spectrum is decreased, or decimated, by a factor equal to the order of smoothing performed.} spectrum to subtract the baseline.  The same baseline-fitting regions and order of polynomial fit were used for both polarizations.   

Flux calibration was derived from simultaneous observations of an internal noise diode.  The baseline-subtracted and flux-calibrated scans for each polarization were accumulated and averaged for each galaxy. A fourth order boxcar smoothing was then applied to the averaged data with decimation, resulting in a resolution of $\sim$6~km~s$^{-1}$ per channel.  The two polarizations were then averaged together to produce the full intensity spectra shown in Figure~\ref{fig:specs}.  We ensured that our flux calibration is correct by comparing the full intensity profile of Mrk~231 with that measured by the VLA \citep{carilli98}; both profiles are identical.  Discussions of individual objects in the Appendix include comparisons with previous data, whose generally good agreement with our data in terms of velocity profile and flux validate our data reduction process.  Table~\ref{tab:sample} also provides a list of some observational parameters for our galaxies.  

Nearly all of the galaxies were detected.  Two galaxies, PG~0804+761 and F15462--0450, had strong RFI spikes overlapping with the expected HI frequencies. At the limit of three hours on source, PG~1126--041 and F21219--1757 were not detected, and F04103--2838 and F15130--1958 were marginally detected.  The detections of PG~1426+015 and PG~1617+175 are uncertain because their putative emission features have widths
similar to those of broadness of the baseline ripples before baseline subtraction.   

\subsection{Spectral Analysis}
\label{sec:simple}

The individual HI profiles in the spectra of our sample galaxy are complex combinations of absorption and emission.  
For simplicity, we have measured the emission and absorption features separately, but we note that this approach may slightly alter the final measured values of these quantities.  More complex decomposition of the emission and absorption is only possible for a few sources because of the degeneracy between emission and absorption, as discussed in Section~\ref{sec:decompose}.

For galaxies with significant emission (peak-to-rms $> 3\sigma$), we measure the central velocity ($V_\odot$), width ($W_{20}$, $W_{50}$), and total velocity-integrated flux ($f_{\rm HI}$) using the GBTIDL task {\it gmeasure}.  In these instances, since there are no obvious companions enclosed within the GBT beam, the total flux is integrated between velocities at which the flux density reaches zero.  The central velocity is the heliocentric velocity measured at the midpoint of the 20\% flux level.  The width of emission is measured at the 20\% and 50\% levels ($W_{20}$ and $W_{50}$, respectively).  The measurement errors in these quantities are derived using the expressions presented in \citet{wei} following the methodologies of \citet{schneider86, schneider90} and \citet{fouque}.  The HI column density is calculated using expression 12.17 from \citet{tools} for optically thin radiation.  Assuming the GBT antenna temperature to flux density gain conversion of $\sim$2~K Jy$^{-1}$: 
\begin{equation}
N_{\rm HI} = 3.64 \times 10^{18} f_{\rm HI} ~{\rm cm}^{-2},
\end{equation}
where $f_{\rm HI}$ is the velocity integrated HI flux in units of Jy~km~s$^{-1}$.  The neutral gas mass is calculated following \citet{haynes}:
\begin{equation}
M_{\rm HI} = 2.35 \times 10^5 f_{\rm HI} (\frac{V_{\rm cor}}{H_0})^2 M_\odot,
\end{equation}
where $V_{cor}$ is the recessional velocity corrected for the influence of the Virgo Cluster, the Great Attractor, and the Shapley Supercluster as calculated by the NASA Extragalactic Database (NED).  All of these measured and derived quantities are listed in Table~\ref{tab:emission}.

Only a few absorption features are significantly detected (by the criterion that peak-to-rms $> 3 \sigma$).  However, many of the absorption features are very broad but shallow, so the peak-to-rms ratio is not necessarily the best determinant of a detection.  We also consider a feature to be detected if its full width at half maximum (FWHM) spans at least 10 velocity channels and it has a peak-to-rms $> 1 \sigma$.  These absorption features are quantified using the GBTIDL task {\it fitgauss}, with one Gaussian component at a time added to a spectrum until the residuals are statistically featureless.  With the exception of PG~1119+120, where a narrow absorption feature is present within the emission profile, only the absorption features are modeled.  For PG~1119+120, the emission was first modeled with Gaussians and subtracted from the data to define a baseline.  The absorption was then modeled from the emission-subtracted spectrum.  The Gaussian models are plotted (as components and as sums) in Figure~\ref{fig:specs}.  The GBTIDL task measures the central velocity ($V_\odot$), FWHM, the ``peak'' flux density, and their measurement errors for each Gaussian component.  The full velocity-integrated absorbed flux ($f_{\rm abs}$), optical depth ($\tau$), and column density ($N_{\rm HI}$) are then calculated from the best-fit models.  We assume the HI absorbing screen covers the nuclear continuum source, in which case the optical depth profile is:
\begin{equation}
\tau(v) = -\ln [1-\frac{S_{\rm HI}(v)}{S_{\rm 1.4~GHz}}],
\end{equation}
where $S_{\rm HI}(v)$ is the HI line profile as a function of velocity and $S_{\rm 1.4~GHz}$ is the continuum flux density \citep{tau}.  The column density is then calculated as:
\begin{equation}
\frac{N_{\rm HI}}{T_{\rm s}} = 1.823 \times 10^{18} \int \tau(v)\,\mathrm{d}v~{\rm cm}^{-2}, 
\end{equation}
where $T_{\rm s}$ is the spin temperature.  In this paper, we will assume a spin temperature of 1000~K (see Section~\ref{sec:mdot}).  For absorption features in PG~1119+120, PG~1244+026, and PG~1440+356, the absorption troughs are deeper than their nuclear continua, suggesting that sources of continuum emission extend beyond the nucleus. Table~\ref{tab:absorption} summarizes the measured and derived values for all of the detected absorption features. 

As a reality check, we inspected the data to confirm that the choice of polynomial order for the baseline fitting does not affect the detection of absorption or emission.  Specifically, we examined whether polynomial orders of 3 or 5 can introduce ripples into the data that masquerade as absorption or emission features.  After repeating the baseline subtraction for our data using a range of polynomial orders (1--5), we confirm that the same spectral features are still detected, even for marginal detections such as that in F04103--2838.  The measured widths and velocities are in agreement with those measured with the best-fit baseline to within the errors.  Therefore, we are confident that even the weak detections noted as "marginal" in Tables~\ref{tab:emission} and \ref{tab:absorption} are real.

\section{HI Profiles of Late-stage Mergers}
\label{sec:profiles}


The individual HI profiles of our targets are discussed in the Appendix.  In general, we see HI trends along the merger sequence but no correlation with infrared luminosity.

All of the detected ULIRGs show HI absorption, typically with multiple velocity components, many of which are narrow (FWHM $\lesssim 100$~km~s$^{-1}$) and some of which are redshifted.  Only 2/7 (29\%) of the detected ULIRGs have HI emission.  The different velocity absorption components have different origins and are not necessarily caused by outflows or infall; kinematics of circumnuclear gas disks may mimic the properties of outflows or infall in single dish data.  In a couple of cases, such as UGC~05101 and Mrk~273, the absorption profiles are nearly symmetrical about zero velocity.  For the Type~2 AGN UGC~05101, the profile is likely from absorption due to a slightly inclined rotating HI disk superposed on an obscured nucleus.  This interpretation is consistent with CO maps of UGC~05101 showing smooth, regular rotation of the molecular gas \citep{genzel98, wilson08}.  On the other hand, the CO velocity field of Mrk~273 is very distorted \citep{downes98, wilson08}, so it is unlikely that the gas is in well-ordered disk-like rotation.  The double absorption in HI seen in Mrk~273 probably corresponds to the binary obscured nuclei in this object.  Thus, CO maps can help eliminate gas disks as the origin of the absorption features.  However, only a few galaxies in our sample have readily available CO maps so direct comparisons are difficult.  Comparison of single dish CO and HI profiles for the QUEST targets is on-going, but it is beyond the scope of the present paper.  The more complex profiles, such as those seen in F15250+3608, F05189--2524, and F15130--1938, may be interpreted as multiple infalling or outflowing components.  They may also result from the superposition of an emission profile and a broad absorption profile.  The latter case is discussed in more detail in Section~\ref{sec:decompose}. 

In contrast, the HI profiles of the FIR-strong quasars are generally detected not only in absorption (88\% or 7 of the 8 detected sources) but also in emission (75\%).  The absorption features are on average $\sim$60~km~s$^{-1}$ broader than those seen in ULIRGs or approximately three times greater than the typical FWHM measurement error in the ULIRGs.  Both redshifted and blueshifted absorption features are seen, although the systemic velocities of these systems are uncertain.  Optical emission lines may underestimate recessional velocity since some fraction of the line emission may arise from outflowing material.  In the case of PG~0050+124 (I Zw 1), this effect causes a shift of $\sim$700~km~s$^{-1}$ (see Appendix).  A similar situation could also hold for PG~1440+356.  In three cases, PG~1119+120, PG~1244+026, and PG~1440+356, the minima of the absorption troughs exceed the nuclear continuum flux densities.  This discrepancy implies that off-nuclear emission, not accounted for in the nuclear fluxes of Table~\ref{tab:sample}, also contributes to the continuum emission in these four sources.  Given resolution limit of the VLA core measurements of PG~1119+120 and PG~1440+356, the off-nuclear emission must be at least $\sim$20--30~kpc from the core; for PG~1244+026, the separation from the core must be at least 3 kpc.  While we cannot formally rule out the possibility that these absorption features are absorption of HI against continuum from background quasars, this possibility is statistically unlikely.  Another possibility is that the flux densities of the cores in these galaxies are time variable as exemplified by the variability of the core in PG~1351+640 (see Appendix).

The HI profiles of the FIR-weak quasars are different from those of the ULIRGs and FIR-strong quasars.  HI is seen mostly in emission (88\% or 7 of 8 detected sources), and only 25\% (2/8) show broad absorption.  The emission profiles are very broad ($W_{50} \gtrsim$500~km~s$^{-1}$) and bumpy, perhaps beginning to resemble those of regularly rotating disks that typically display Gaussian or double-horned profiles; the measured widths and velocities are consistent with previous measurements of quasars \citep[e.g.,][]{bothun84, condon85, hutchings87, ho08, konig09}.  These properties will be revisited in Section~\ref{sec:decompose}.

\begin{figure}
\centering
\includegraphics[width=3.5in]{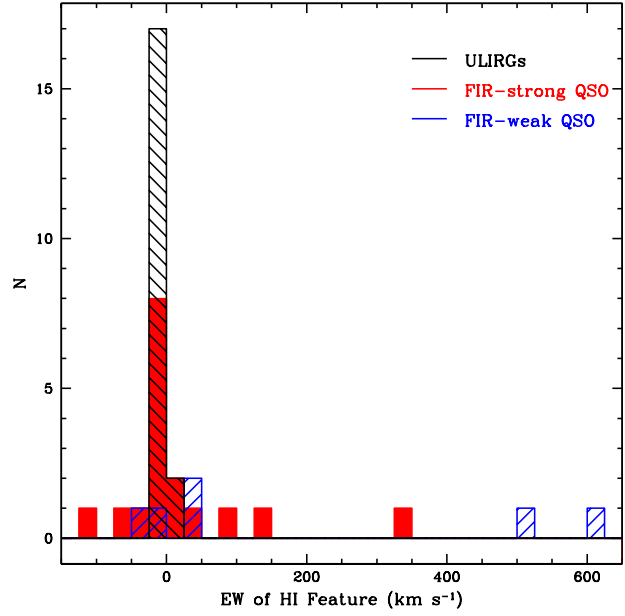}
\caption{The distribution of equivalent widths of the emission and absorption components as listed in Tables~\ref{tab:emission} and \ref{tab:absorption}.  The absorption components are shown as having negative EWs.  Targets with only upper limits to their continuum flux densities are excluded from this plot.  Along the merger sequence, the ULIRGs have the weakest features in both emission and absorption, the FIR-strong quasars have the strongest absorption features with moderate strength emission features, and the FIR-weak quasars are dominated by the strongest emission features.   }
\label{fig:ew_dist}
\end{figure}

\begin{figure}
\centering
\includegraphics[width=3.5in]{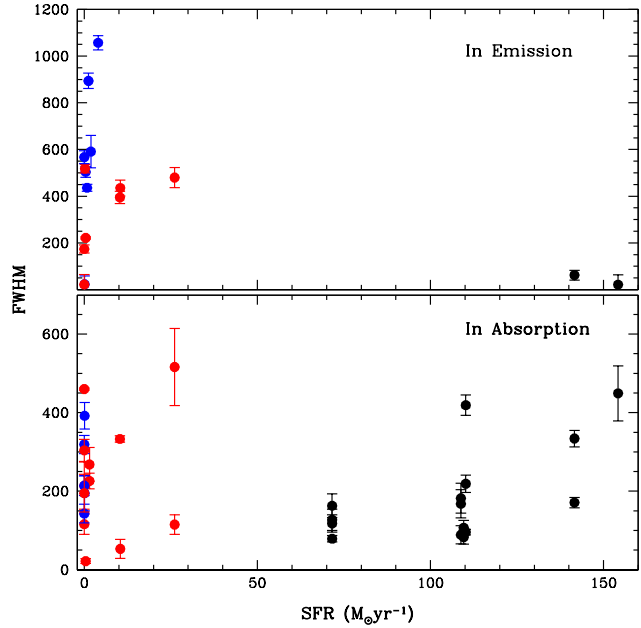}
\caption{The velocity widths of the components measured in emission (top; $W_{50}$ in Table~\ref{tab:emission}) and absorption (bottom; FWHM in Table~\ref{tab:absorption}).  The colors represent ULIRGs (black), FIR-strong quasars (red), and FIR-weak quasars (blue).  The width of the HI emission in the top panel may be inversely correlated with the SFR.  The galaxies with the weakest star formation (FIR-weak quasars) have the strongest emission features, while the galaxies with the strongest star formation (ULIRGs) have the narrowest velocity widths.  The velocity widths of the FIR-strong quasars fall between those of the ULIRGs and the FIR-weak quasars.  There also seems to be a weak correlation between the FWHM in absorption and the SFR for ULIRGs, but this result is very uncertain as ULIRGs tend to have multiple absorption components for a single SFR.  There is no apparent correlation between the SFR and the absorption parameters in the quasars, suggesting that the outflow-related absorption components are not the result of star formation. }
\label{fig:fwhmvsfr}
\end{figure}

Considering now our entire HI sample, the distribution of equivalent widths (EW; defined as the integrated flux density for each component divided by the continuum flux density of the core) is plotted in Figure~\ref{fig:ew_dist}.  For this definition, the equivalent widths of the absorption components are negative.  It is clear that the strengths of the emission and absorption features progress along the merger sequence, with ULIRGs having the weakest features, the FIR-strong quasars having the strongest absorption features, and the FIR-weak quasars having the strongest emission features.  In Figure~\ref{fig:fwhmvsfr}, we compare the FWHMs of emission and absorption with the star formation rates of their host galaxies.  For emission, the FIR-weak sources have the broadest widths, the ULIRGs have the narrowest widths, and the FIR-strong quasars bridge the gap between those two classes.  In the ULIRGs, there appears to be a weak correlation between the width of the absorption components and the SFR, reminiscent of a similar relationship seen in Na~I \citep{rupke05b}.  However, given that multiple absorption components are associated with a single system (and thus a single SFR value), the validity of the connection between the SFR and the FWHM of the absorption components in ULIRGs is uncertain.  On the other hand, absorption FWHM is not at all related to SFR in the quasars, further suggesting that the AGN play a role in these systems. 

\section{Polarization-dependent HI Absorption}
\label{sec:polar}


An unexpected outcome of our GBT program is that many of the absorption features show a dependence on polarization (Figure~\ref{fig:polspecs}).  Indeed, 13 of our 23 GBT-detected targets show some profile difference in their two polarizations (1 ULIRG, 7 IR-excess quasars, and 5 optical quasars).  For GBT observations, it is expected that there is a $\lesssim$6\% difference between polarizations in terms of normalization, but the observed spectral shapes should be the same in both polarizations (R. Maddalena, private comm.).  This expectation is confirmed by the fact that our M82 data remain in agreement after baseline subtraction.  Polarization variation in our QUEST sample is not RFI-related, as we have removed time intervals where the data are affected by broadband RFI.  Since it is improbable that the HI screen itself is polarized, the underlying continuum emission must be polarized.  

\subsection{Time Variability}
\label{sec:timevar}

\begin{figure}
\centering
\includegraphics[width=2.3in, angle=-90]{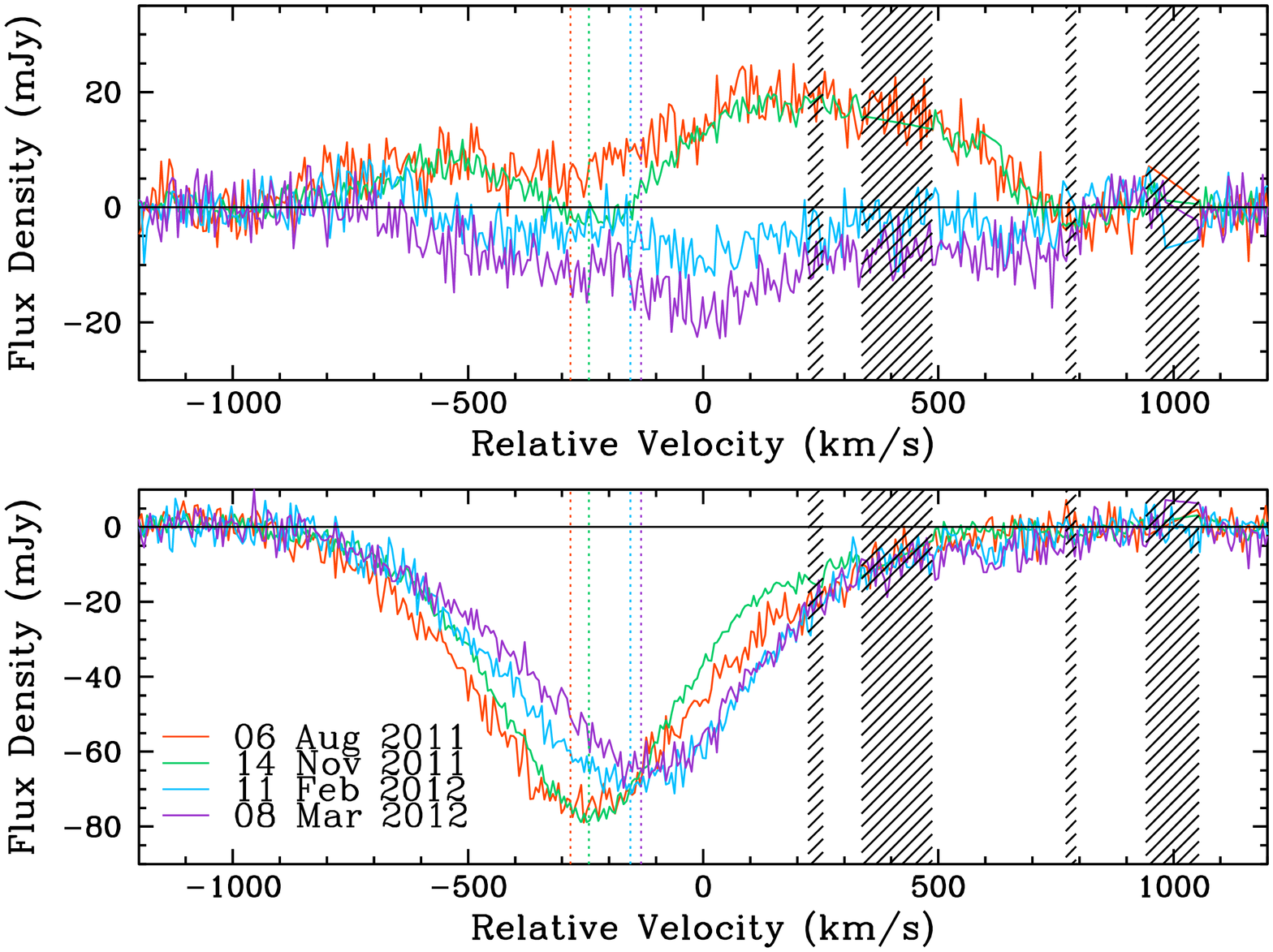}
\includegraphics[width=2.3in, angle=-90]{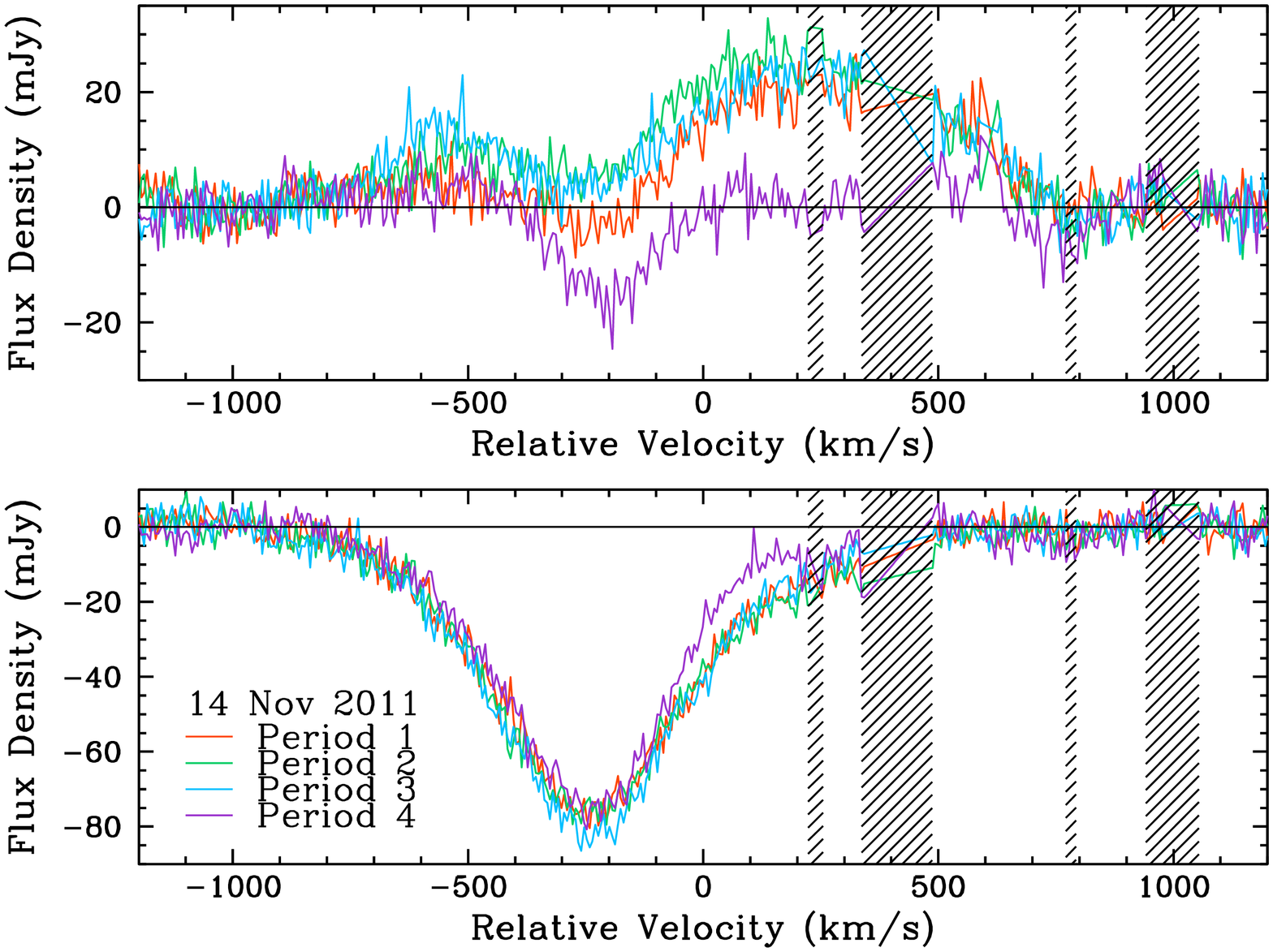}
\caption{{\it Left:} GBT spectra of PG~1211+143 obtained at four epochs separated by at least one month, with the YY spectrum above and the XX spectrum below.  The shifts in the central velocity of the deep absorption component are denoted by the vertical dotted lines in their corresponding colors.  The central velocity shifted by $\sim$120~km~s$^{-1}$ over the seven-month period covered by our observations.  {\it Right:}  The 2011 November spectrum broken down into four 30-minute periods.  There is clear variability in the shape of the spectrum, suggesting a varying continuum against which the absorption is detected.  If the variability is real, it is consistent with the jet-induced outflow picture suggested by \citet{morganti11}.}
\label{fig:pg1211var}
\end{figure}

\begin{figure}
\centering
\includegraphics[width=2.3in, angle=-90]{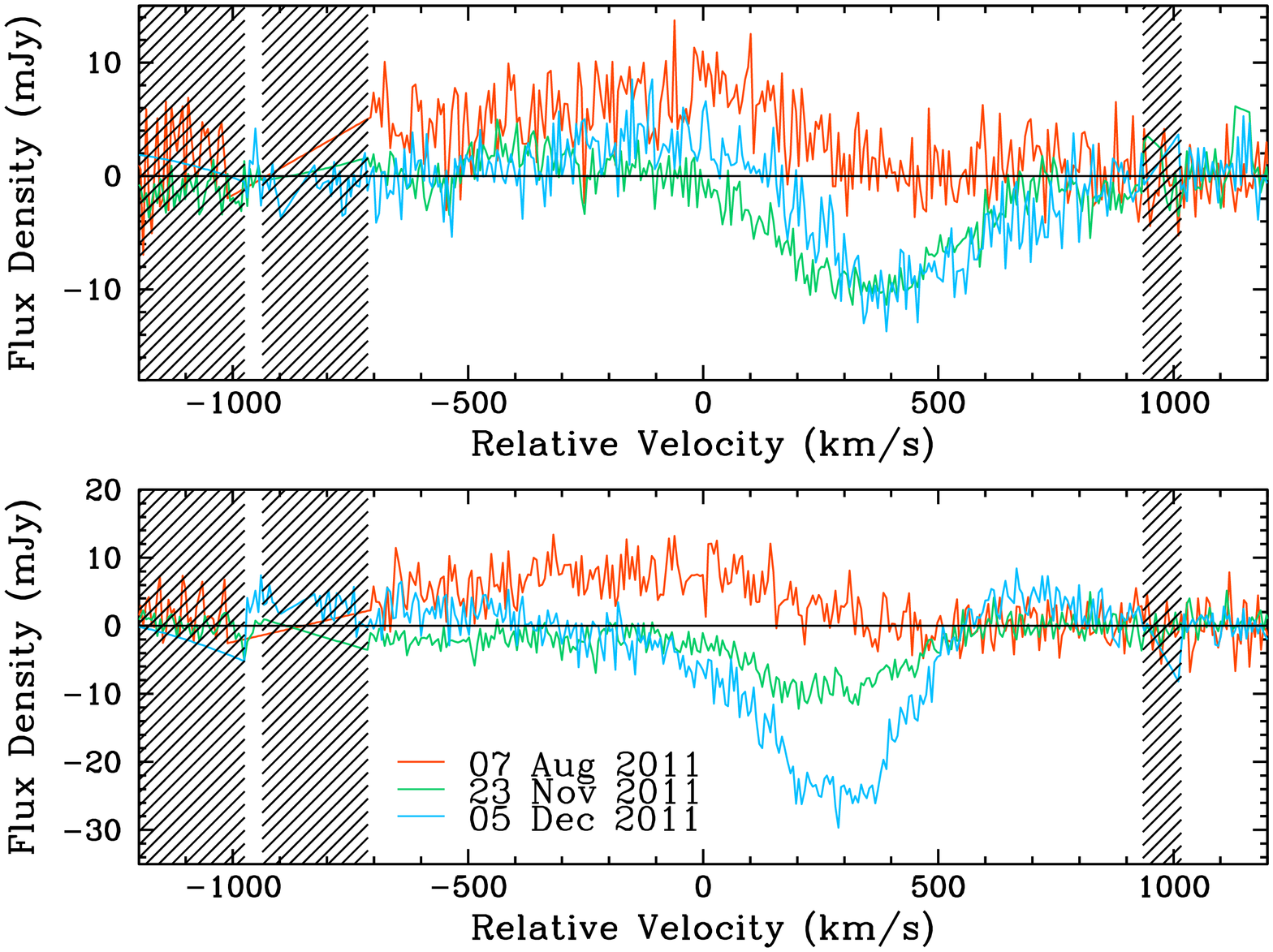}
\includegraphics[width=2.3in, angle=-90]{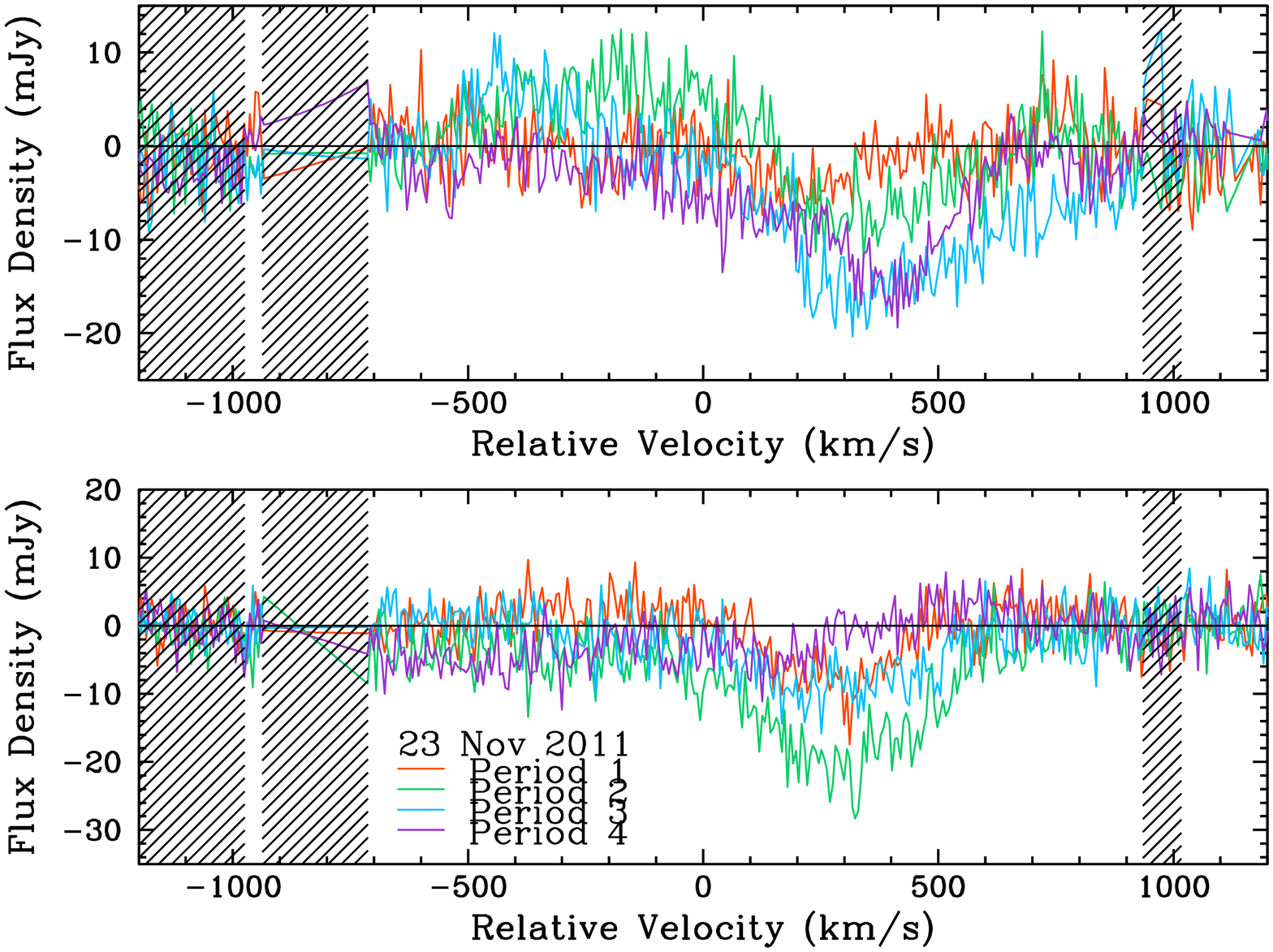}
\caption{Same as Figure~\ref{fig:pg1211var}, but for PG~1440+356.  Short timescale variations are seen in this source, similar to those in PG~1211+143.}
\label{fig:pg1440var}
\end{figure}

The most dramatic examples of polarized spectra are seen in PG~1211+143 and PG~1440+356.  Not only are these spectra polarized, but they are also time variable on timescales of $\sim$30 minutes (Figures~\ref{fig:pg1211var} and \ref{fig:pg1440var}).  In fact, of our radio-detected sample galaxies, time variability is only seen in sources that have significant absorption and spectra that are significantly polarized.  A possible concern is that since we observed these sources over several sessions, variability could be an instrumental effect.  However, comparing the different sessions' data for M82, we find that time variability and polarization are not seen beyond what is expected of typical GBT observations.  Moreover, session-to-session instrumental variability would not explain the 30-minute variability seen within the same observing session.  Time varying polarizations are not seen only in the targets with the brightest continua, so the variations are not likely to be due to instrumental systematics from poor cancellation of baseline ripples in the position-switching scans \citep{oneil02}.  

A different sort of instrumental effect may be leakage of linear polarization from the linearly polarized background continuum into the total intensity of the absorption profiles.  We cannot yet exclude this scenario since these observations were not designed for polarimetry and our current data lack the cross-correlated products (XY and YX) that would be needed to confirm these time variations.

\begin{figure}
\centering
\includegraphics[width=2.3in, angle=-90]{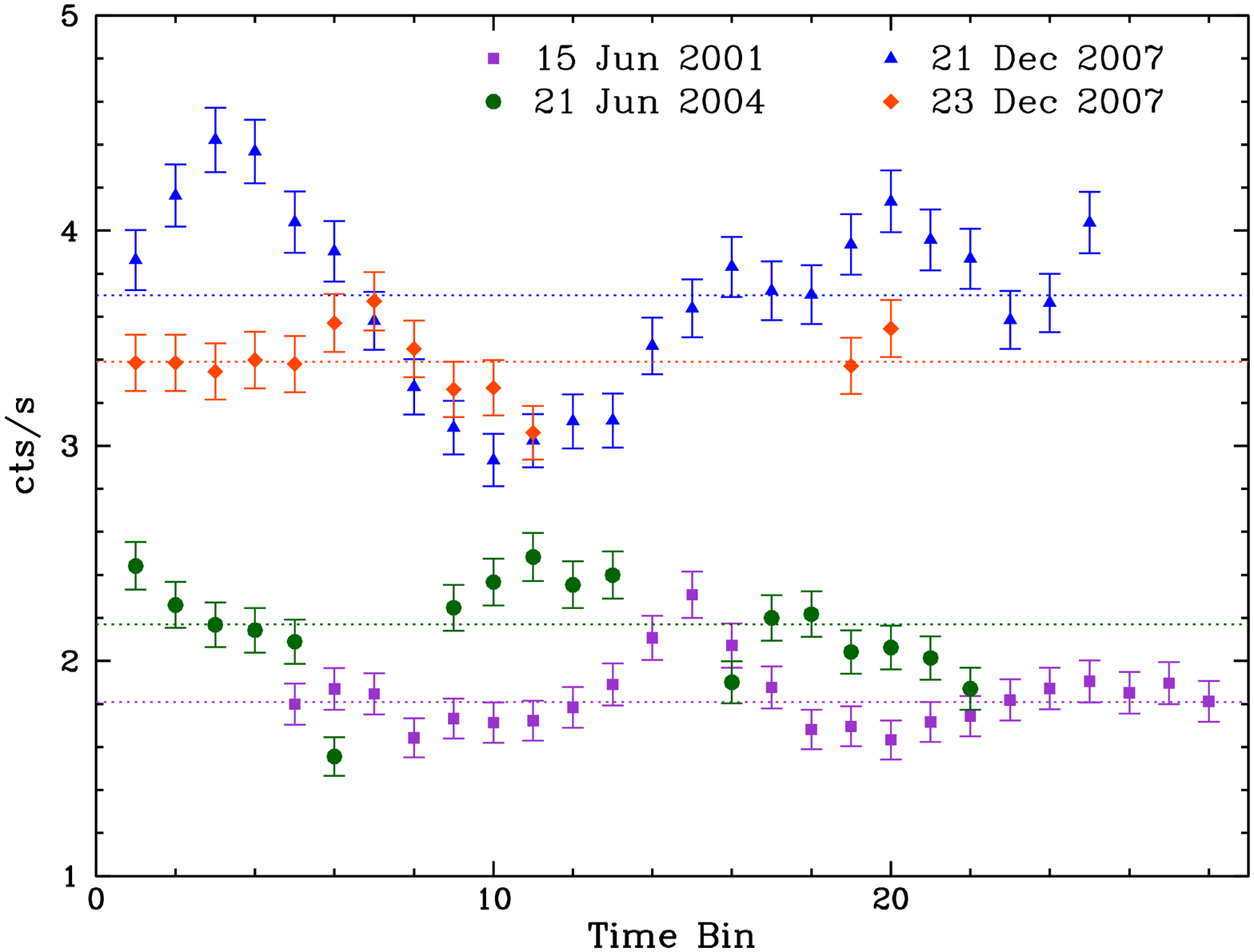}
\includegraphics[width=2.3in, angle=-90]{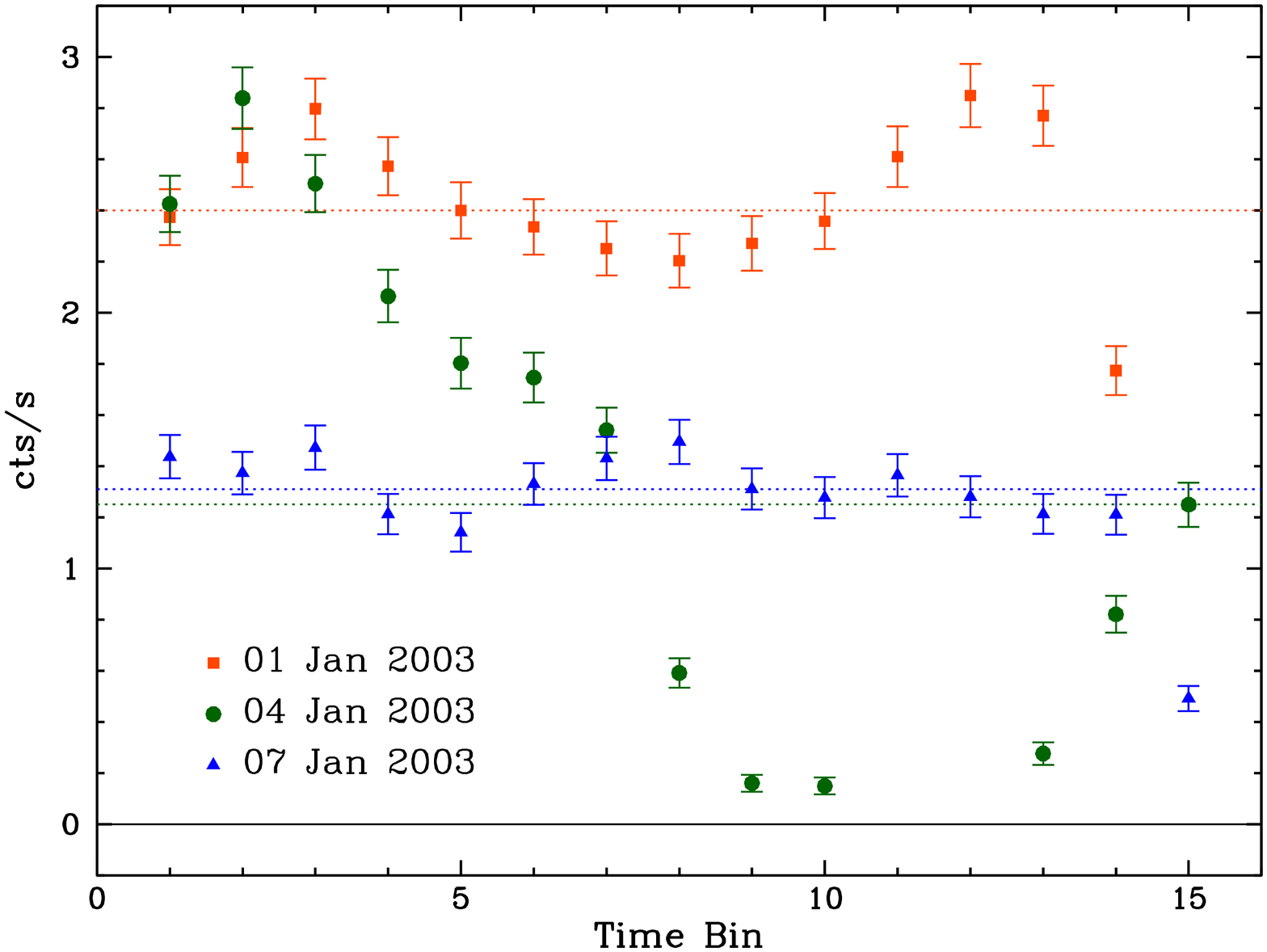}
\caption{Background-subtracted 0.5--10~keV X-ray light curves of PG~1211+143 (left panel) over 6.5 years through four separate observations, and of PG~1440+356 (right panel) over one week through three separate pointings, from {\it XMM-Newton}.  For each epoch, each point represents the EPIC-pn count rate in 1800~s intervals starting with the beginning of the observation in the first bin.  The dotted lines are the median rates for each specific epoch.  The periods of high background flares have been removed from the plot and the calculations of the median rate.  The error bars are 3$-\sigma$.  The AGN have both high and low flux states and are clearly variable on the 30-minute time scale seen in Figures~\ref{fig:pg1211var} and \ref{fig:pg1440var}.}
\label{fig:xrayvar}
\end{figure}

If these time variations are real, they could be caused by one of several processes.  First, they could be due to variations in the compact continuum sources behind the HI screen.  As the X-ray light curves of PG~1211+143 and PG~1440+356 show (Figure~\ref{fig:xrayvar}), the X-ray emitting central region can be highly variable on a timescale of about 30 minutes.  Although not contemporaneous with our GBT data, these X-ray data suggest that the AGN continuum can vary on a 30-minute time scale if the radio continuum is indeed related to the X-ray producing regions.  However, if the polarization-dependent variations are simply a result of the changing compact source continuum, then we would expect the continuum to vary in block and would observe a change in the equivalent width of the HI absorption, not the frequency-dependent variations we see in these targets.  Second, variability could be a result of changes in the structure of the HI screen.  However, the timescale for the destruction of small HI clouds in the interstellar medium is on the order of a Myr, depending on the exact properties of the interstellar medium causing the absorption \citep[e.g.,][]{mc77, nik06}; it is therefore unlikely that cloud destruction causes the observed short-term variations.  Similarly, it is unlikely that the movement of HI clouds into and out of our line of sight to the continuum source can cause the very broad variations observed in these objects. 

\begin{figure}
\centering
\includegraphics[width=4in]{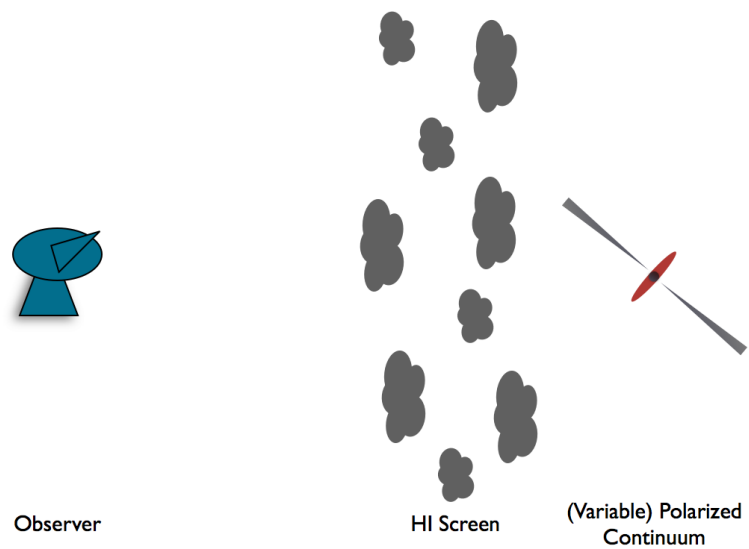}
\caption{The HI screen and source configuration that explains the variable polarization-dependent spectra observed in PG~1211+143 and PG~1440+356.  In this picture, the HI screen is extended, covering both the core and the radio jets.  The jets produce the polarized continuum against which the absorption features are detected.  As the central source fluctuates, the ratio of central-to-jet emission also changes, resulting in the observed spectra. }
\label{fig:gbtlos}
\end{figure}

The time variability of the polarization-dependent HI absorption is easier to explain if the polarized continuum emission comes from both the nuclear source and an extended source (e.g., radio jets).  In this picture (Figure~\ref{fig:gbtlos}), the HI screen is also extended, covering both the nuclear source and extended jets with kinematically distinct clouds.  As the central source fluctuates, the ratio of central-to-jet emission also changes, resulting in the observed wavelength-dependent polarization variations.  In this scenario, there is no need for the structure of the HI screen to change on a fast timescale.



\subsection{Decomposing the Spectrum}
\label{sec:decompose}

Regardless of whether the time variability is real, it is certainly the case that the spectra shown in Figure~\ref{fig:polspecs} are polarized.  Again, we checked for baseline-subtraction-related effects on the detection of polarized absorption and found that the choice of the polynomial order of the baseline fit matters little to the detection of these polarization-dependent absorption features.  The most obvious example of polarization not due to observational systematics is seen in the spectra of PG~0050+124 (I Zw 1).  This HI profile is not polarized at relative velocities of 500--1000~km~s$^{-1}$ but becomes polarization-dependent between relative velocities --500 and +500~km~s$^{-1}$ (Figure~\ref{fig:polspecs}).  This behavior argues against attribution of apparent polarization dependencies to poor baseline subtraction or session-to-session variations in our GBT observations.  At least in this one case, the polarization dependence must be due to something physically happening at relative velocities between --500 and +500~km~s$^{-1}$.  When we compare the HI profiles of each galaxy in the two separate polarizations, it appears that one polarization typically has higher flux than the other, with the exception of PG~1440+356, in which there is a velocity shift instead.  The effect is not systematic in the sense that one polarization is always higher than the other; each is nearly equally likely to be the polarization with the higher intensity.  In some cases, particularly in PG~1351+640 and PG~1411+442, the lower-intensity profile appears to be the same as the higher-intensity profile with the addition of a broad absorption component.  

To follow up on this intriguing hint, we produced the difference spectra (the spectrum in the polarization with the more negative HI profile minus the spectrum in the polarization with the more positive HI profile) shown in Figure~\ref{fig:diffspecs}.  The absorption features in these spectra are then modeled the same way as before, using the GBTIDL task {\it fitgauss}.  When these absorption models are included, many of the full intensity spectra that need multiple components to explain the blue- and red-shifted components can be explained simply by HI emission that is combined with a single broad absorption component.  This pattern is most obvious in PG~1351+640 and PG~1411+442.  The results of this type of complex decomposition of the spectra are provided in Tables~\ref{tab:complex_emission} and \ref{tab:complex_absorption}.  Similar decompositions can explain the shapes of the full intensity spectra of the ULIRGs F05189--2524 and F15130--1958, and the FIR-strong quasar PG~1244+026.  However, since those spectra are not polarized, we cannot create difference spectra to model the broad absorption component.  The {\it fitgauss} task statistically prefers the multiple-component models to the single-absorption model for these spectra.  Overall, with the exception of PG~1501+106, the decompositions result in stronger features in both absorption and emission.  

\begin{figure}
\centering
\includegraphics[width=3.5in]{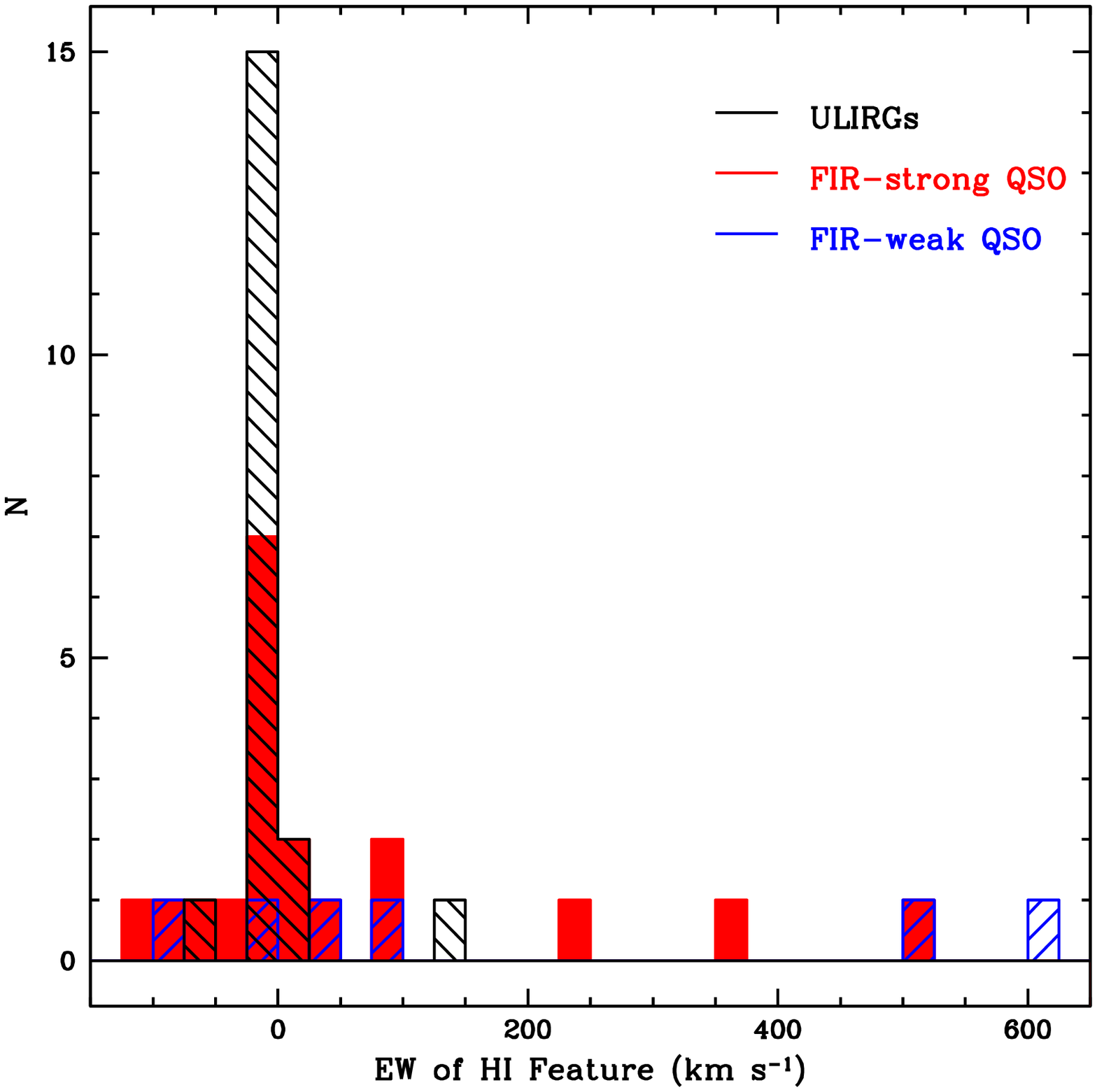}
\caption{Same as Figure~\ref{fig:ew_dist} apart from the use of the complex fitting results when available (Tables~\ref{tab:complex_emission} and \ref{tab:complex_absorption}).   The general trends with stage along the merger sequence are in agreement with those in Figure~\ref{fig:ew_dist}.}
\label{fig:ew_dist_com}
\end{figure}

\begin{figure}
\centering
\includegraphics[width=3.5in]{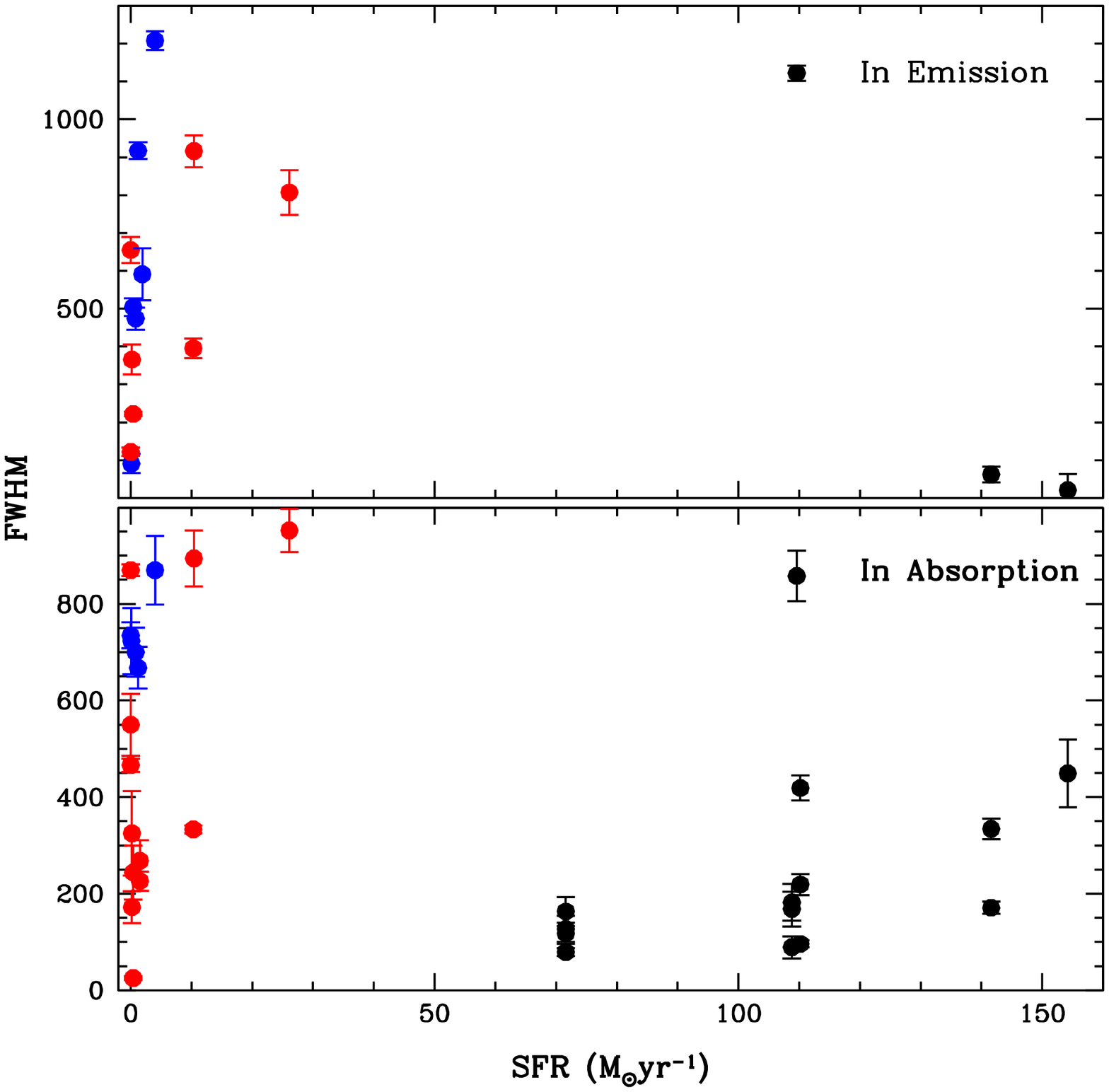}
\caption{Same as Figure~\ref{fig:fwhmvsfr} apart from the use of the complex fitting results in Section~\ref{sec:decompose} when available (Tables~\ref{tab:complex_emission} and \ref{tab:complex_absorption}).   The general trends are in agreement with those in Figure~\ref{fig:fwhmvsfr}.}
\label{fig:fwhmvsfr_com}
\end{figure}

Given the results of this exercise, we find the instances of HI absorption (emission) to be 100\% (29\%) in detected ULIRGs, 100\% (88\%) in FIR-strong quasars, and 63\% (100\%) in FIR-weak quasars.  The more complex decomposition of the spectra does not drastically change the distribution of EWs (Figure~\ref{fig:ew_dist_com}) compared to Figure~\ref{fig:ew_dist}.  Figure~\ref{fig:fwhmvsfr_com} provides a comparison of the FWHMs of the emission and absorption features from the complex modeling with the star formation rates of these galaxies.  As in Figure~\ref{fig:fwhmvsfr}, the FIR-weak quasars have the broadest emission features and the ULIRGs have the narrowest emission features.  In terms of absorption, there is no obvious relationship between the absorption feature widths and the star formation rates.  As before, the FIR-weak quasars have the broadest absorption features, with the FIR-strong quasars spanning the range of widths between FIR-weak quasars and ULIRGs.

\section{The Properties of HI Outflows}
\label{sec:origin}

As discussed in Section~\ref{sec:profiles}, the absorption components are not necessarily caused by infall or outflow.  It is possible, as seen in UGC~05101 (edge on) and Mrk~231 (face on), that these features can be associated with the kinematics of gas disks.  A circumnuclear gas disk seen nearly edge on with a non-isotropic gas mass (or background continuum) distribution can mimic characteristics of infall and outflow.  Unfortunately, our single dish data cannot unambiguously exclude gas disks as the source of the features we observe, but it is unlikely that all of our targets with absorption features have gas disks that are seen precisely edge-on. 

Keeping this caveat in mind, we assume in the following discussion an outflow origin only for blueshifted HI absorption features with HI-corrected heliocentric velocities ($V_{\odot,~{\rm HI}}$\footnote{These systemic velocities are relative to the HI emission rather than the optical emission.  See Table~\ref{tab:outflows} for more details.  A remaining source of uncertainty in the interpretation of HI line profiles is the question of whether the midpoint of the 20\% flux level is always a good approximation of the true systemic velocity.  Emission from molecular lines is a more reliable tracer of systemic velocity than HI, so the ideal solution here would be acquisition of single-dish CO spectroscopy for the full QUEST sample.  Unfortunately, these observations are still pending.}) below --50~km~s$^{-1}$ \citep{rupke05a}.  A number of mechanisms have been proposed to explain such HI outflows.  They include adiabatically expanding broad emission line clouds \citep[e.g.,][]{elvis02}, starburst- and AGN-driven galactic winds \citep[e.g.,][]{heckman90, vei05}, and AGN-driven jets interacting with the surrounding interstellar medium \citep[ISM; e.g.,][]{oosterloo00}.  These mechanisms predict different outflow locations (from parsec to kiloparsec scales) and geometries (wide angled vs. collimated).  

\subsection{Mass Outflow Rates}
\label{sec:mdot}

Following \citet{heckman00} and \citet{rupke02}, the mass outflowing ($\dot{M}$) at a velocity $v$ is calculated using: 

\begin{equation}
\frac{\dot{M}}{M_\odot~{\rm yr^{-1}}} = 20 (\frac{\Omega}{4 \pi})(\frac{r_*}{\rm 1~kpc})(\frac{N_{\rm HI}}{\rm 10^{21}~cm^{-2}})(\frac{v}{\rm 200~km~s^{-1}}), 
\end{equation}
assuming spherical geometry, mass conservation, and a wind that flows from a distance of $r_*$ to infinity.  The outflow is also assumed to subtend a solid angle of $\Omega$ within which the gas has a covering fraction of 1.  To facilitate comparisons with the values derived from young radio galaxies by \citet{morganti05b}, we use the same assumptions of $\Omega = \pi$ and $r_* = 1$~kpc.  Applying these assumptions and Equation~5 to the blueshifted absorption components where $v = V_{\odot,\rm ~HI}$, we find mass outflow rates $\dot{M}$~(wide) up to more than 6000~$M_\odot$~yr$^{-1}$ (summarized in Table~\ref{tab:outflows}).  These values are much larger than those of $\lesssim$60~$M_\odot$~yr$^{-1}$ measured by \citet{morganti05b} and the range measured in ULIRGs \citep[13--133~$M_\odot$~yr$^{-1}$;][]{rupke02}.

Many of the mass outflow rates we infer seem unrealistically large; such high rates have not been measured in other astrophysical objects.  These values do, however, depend on the assumptions we have made above.  The mass outflow rate depends linearly on the size scale of the outflow region and the derived HI column density.  Recall from Equation~4 that the column density depends on the spin temperature and optical depth profile, which in turn depends on the brightness of the background continuum.  There are thus three key factors that affect our estimates of a mass outflow rate:  the spin temperature of the gas, the assumed distance, and the background continuum.  The assumption of 1000~K for the HI spin temperature seems to be reasonably conservative.  In the vicinity of powerful AGN, the extreme conditions of the gas and the presence of shocks can result in spin temperatures $>$3000~K \citep[e.g.,][]{holt06}, although we note that absorption-line opacity will be dominated by the coldest material in any multiphase medium.  Spatially resolved interferometric maps of several quasars suggest that the outflow regions have size scales $\sim$0.2--1.6~kpc \citep{oosterloo00, morganti03, emonts05, morganti05a}.  
Adopting the smaller distance of 0.2~kpc instead of 1~kpc would reduce the derived mass outflow rates by a factor of 5, more consistent with those derived for ULIRGs.


Finally, the estimated HI column density also depends strongly on the geometry of the source of continuum emission.  We have assumed that this emission comes entirely from the core.  However, the absorption troughs in several quasars in our sample are deeper than the core fluxes.  This excess implies that off-nuclear emission also contributes to the observed continuum.  In this case, $\tau$ in Equation~(3) is overestimated and so is $N_{\rm HI}$, by a factor equal to the actual nuclear/total continuum flux ratio, although a larger source size may compensate to some extent.


\subsection{Jet-induced Outflows}
\label{sec:jet}

If the off-nuclear continuum source is indeed physically related to the target and not a background quasar that happens to lie within the GBT beam, our polarized spectra give more credence to the idea that radio jets play an important role in driving HI gas outflows \citep[e.g.,][]{morganti11}.  Several sources (see Appendix) in our sample with polarized spectra have spatially resolved radio jets detected at higher frequencies.  The resolutions of these VLA data imply that the jets are launched within $\sim$0.4--0.9~kpc of the core.  Considering a jet-induced outflow scenario for the HI absorption features, we again apply Equation~5 to the blueshifted outflow components and assume $r_* = 0.5$~kpc and a narrow opening angle of 5$^\circ$ for the jet.  Under these new assumptions, the range of mass outflow rates is 0.2--31~$M_\odot$~yr$^{-1}$ (see col. (8) of Table~\ref{tab:outflows}).  We consider these values lower bounds on the actual mass outflow rates, while those derived using $\Omega = \pi$ and $r_* = 1$~kpc (col. (7) of Table~\ref{tab:outflows}) are upper bounds.

Under the jet scenario, the kinetic energy release rate can be calculated using the equation from \citet{heckman00}:
\begin{equation}
\frac{\dot{E}}{\rm erg~s^{-1}} = 2.7 \times 10^{41} (\frac{\Omega}{4 \pi}) (\frac{r_*}{\rm 1~kpc})(\frac{N_{HI}}{\rm 10^{21}~cm^{-2}})(\frac{v}{\rm 200~km~s^{-1}}).
\end{equation}
The kinetic energy in the outflow is on the order of $10^{38}$ to $10^{42}$ erg~s$^{-1}$ for our sample.  

The association of jets with the detected HI absorption features is also seen in the distribution of radio loudness of our sample.  Only one quasar in our sample is considered radio-loud based on optical-to-radio flux ratios (see Appendix), but optical data may be subject to contamination from the host galaxy, resulting in an underestimate of radio loudness.  Using the 1.4~GHz to 2--10~keV ratio as an indicator of radio loudness and a radio-loud cutoff value empirically derived from 1600 AGN \citep{lafranca}, we find that many objects in our sample are radio-loud (Table~\ref{tab:sample}).  These results are consistent with those found in a recent study by \citet{chandola} that suggest higher detection rates of HI absorption toward extended compact steep-spectrum and gigahertz-peaked spectrum sources than toward the cores of radio galaxies.  The environments in which the jets interact with the interstellar medium can be complex, and we may be seeing a strongly polarized continuum being absorbed selectively by multiple neutral hydrogen clouds, resulting in polarized HI absorption.  

The fraction of radio-loud sources with associated HI absorption (58\%) is far higher than the $\sim$10\% observed in the general population of AGN.  Given the high star formation rates of the ULIRGs, the circumnuclear starburst can contaminate the 1.4~GHz measurements. Using the 1.4~GHz to SFR correlation of \citet{bell03}, we corrected the 1.4~GHz flux for star formation.  Of the three ULIRGs with blueshifted absorption and X-ray detections, all (100\%) are radio-loud.  Similarly, of the six FIR-strong quasars with detected HI outflows and X-ray emission, three (50\%) are considered radio-loud; one of the two (50\%) FIR-weak quasars with HI outflows and X-ray flux are radio-loud.  This pattern recalls an idea put forward by \citet{wilson95} on the formation of radio galaxies to explain why radio-quiet AGN outnumber radio-loud AGN.  The authors suggested that major mergers, such as the progenitors of our sample of galaxies, result in coalesced supermassive black holes with high spin values due to the conservation of angular momentum.  Substantial energy is then extracted from these fast-spinning black holes to drive jets, yielding radio-loud AGN.  Thus, younger radio galaxies may have more powerful jets and therefore stronger jet-induced outflows.  This scenario is consistent with the observations of \citet{morganti05b}, whose sample of young or recently restarted radio galaxies show similar HI outflows.  The \citet{wilson95} picture is also consistent with theoretical models of \citet{tchek10} that suggest the radio loud/quiet dichotomy in AGN is due to the spin of the central black hole.

\subsection{HI Outflows Along the Merger Sequence}
\label{sec:evolve}

\begin{figure}
\centering
\includegraphics[width=3in]{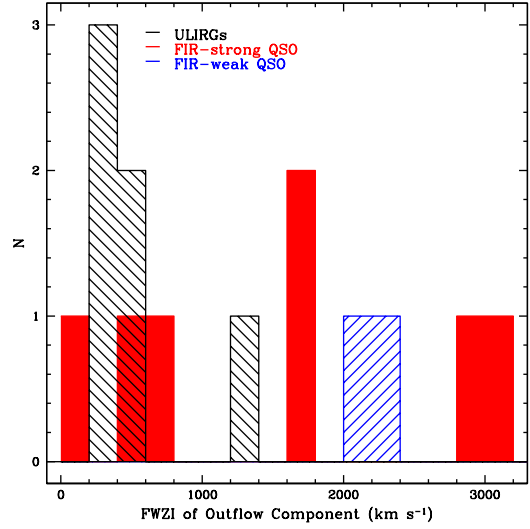}
\caption{The distribution of FWZI of the absorption components in galaxies with detected outflows with $V_{\odot,~{\rm HI}}$ less than --50~km~s$^{-1}$ (see Table~\ref{tab:outflows}).  There is a tentative progression of velocity widths from ULIRGs to FIR-strong quasars to FIR-weak quasars.}
\label{fig:fwzidist}
\end{figure}

Even though our sample with mass outflow rates is relatively small, we see some trends in our outflow measurements (Table~\ref{tab:outflows} and Figure~\ref{fig:fwzidist}).  However, given the small number of detected HI outflows, particularly with only two measurements for the FIR-weak quasars, we cannot statically confirm that the distributions we see in Figure~\ref{fig:fwzidist} is significant.  There is also no obvious relationships between mass outflow rate and the bolometric luminosity, AGN luminosity, or radio loudness for our sample.  

Systems with deep absorption and outflows become more prevalent as we progress along the merger sequence from the newly coalesced ULIRGs to the FIR-strong quasars.  The trend between outflow kinematics and star formation rates among the ULIRGs (Figures~\ref{fig:fwhmvsfr} and \ref{fig:fwhmvsfr_com}) suggests that the HI outflows in these objects are driven mainly by the starbursts, with a median mass outflow rate in the jet scenario ($\dot{M}$ (jet) $\sim$1~$M_\odot$~yr$^{-1}$) similar to their SFRs and previous Na~I-based outflow measurements \citep{rupke02, rupke05a, rupke05b, rupke05c, martin05, martin06}.  ULIRGs also tend to have the narrowest absorption features.  
In the FIR-strong quasar systems, the median mass outflow rate is $\dot{M}$ (jet) $\sim$5.8~$M_\odot$~yr$^{-1}$, considerably larger than their median SFR of $\sim$1.0~$M_\odot$~yr$^{-1}$.  
While the median mass outflow rate of the FIR-weak quasars is lower, $\dot{M}$ (jet) $\sim$0.3~$M_\odot$~yr$^{-1}$, so is their median SFR ($\sim$0.6~$M_\odot$~yr$^{-1}$).  Thus, the mass-loading factor, {\em i.e.}, the ratio of mass outflow rate to SFR, is much greater in the FIR-strong than the value of near unity derived in ULIRGs.  The same value is much lower in the FIR-weak quasars, but this is difficult to compare given that this is based on a single measurement.  However, the distinction between the ULIRGs and FIR-strong quasars further supports the idea that the AGN is playing a role in driving HI outflows in the quasars before the end of large-scale star formation.  

HI in emission is much more prevalent among the quasars than the ULIRGs, and the HI emission profiles increasingly resemble those of regularly-rotating neutral gas disks as we move from FIR-strong to FIR-weak quasars.  The amount of HI seen in emission increases modestly from $\sim$4$\times$10$^{10}$~$M_\odot$ among the FIR-strong quasars to $\sim$7$\times$10$^{10}$~$M_\odot$ among the FIR-weak quasars.  These masses are consistent with those measured in low-redshift quasars \citep{konig09} and the high end of the gas mass range seen in nearby active galaxies \citep{ho08}.  We may be witnessing the creation of HI disks from residual neutral gas at the centers of merger remnants.  

\section{Summary}
\label{sec:sum}

We have analyzed of GBT HI spectra of 27 ULIRGs and quasars selected from the QUEST sample.  We find that:

\begin{enumerate}

\item As mergers evolve from fully coalesced ULIRGs to FIR-strong quasars to FIR-weak quasars, the characteristics of their HI profiles also change from mostly absorption in the ULIRGs to mostly emission in the FIR-weak quasars, with the FIR-bright quasars having intermediate properties.  The kinematics of these systems change from narrow absorption features likely due to outflows driven by starbursts in ULIRGs, to broad absorption features from jet-driven outflows driven by AGN in FIR-strong quasars, and finally to the emission profiles more characteristic of gas in rotation in FIR-weak quasars.

\item A large fraction of the quasars exhibit polarization-dependent absorption features, with 78\% of FIR-strong quasars and 63\% of FIR-weak quasars showing these features compared with 13\% in ULIRGs.  Some of these features vary on short time scales (down to $\sim$30 minutes).  Neither time variability nor polarization-dependent spectral features are seen in contemporaneous observations of M82 (as expected), reducing the likelihood that these effects are instrumental artifacts.  These features are likely due to HI absorption of polarized nuclear and off-nuclear continuum emission.  The large fraction ($\sim$60\%) of radio-loud sources with HI outflow-associated absorption suggests a close relationship between radio jets and the HI screen.

\item The polarization-dependent absorption features are most likely due to the emergence of radio jets in these systems, contributing to the continuum emission against which the absorption features are detected.  The $\sim$60\% radio-loud fraction of our sources with HI outflows is far higher than the $\sim$10\% seen in the general population of AGN.  This distinction suggests that radio jets play an important role in shaping the environments in these systems, which may tract a transition phase in the evolution of gas-rich mergers into ``mature'' radio galaxies.

\end{enumerate}

\acknowledgements

We are grateful to the anonymous referee for providing useful comments that improved the paper.  We thank the GBT operators and the Green Bank staff, especially Ron Maddalena, for support during this program (11B-075, 12A-123), as well as Tim Robishaw and Lisa Wei for useful discussions.  We also thank Sheila Kannappan and her team for allowing us to examine their proprietary data on PG~1501+106.  The National Radio Astronomy Observatory is a facility of the National Science Foundation operated under cooperative agreement by Associated Universities, Inc. We made use of the NASA/IPAC Extragalactic Databased (NED), which is operated by the Jet Propulsion Laboratory, Caltech, under contract with NASA.  We also acknowledge the usage of the HyperLeda database (http://leda.univ-lyon1.fr).  S.H.T. is supported by a NASA Postdoctoral Program (NPP) Fellowship.  S.V. acknowledges support from a Senior NPP award held at the NASA Goddard Space Flight Center and from the Humboldt Foundation which provided funds for a long-term visit at MPE in 2012.

{\it Facilities:} \facility{GBT}.


\appendix
\section{Notes on Individual Galaxies}

In this appendix, we describe the HI spectral properties of our GBT sample.  When appropriate, we compare the present data with previous observations of these galaxies.

\begin{figure}
\centering
\includegraphics[width=7in]{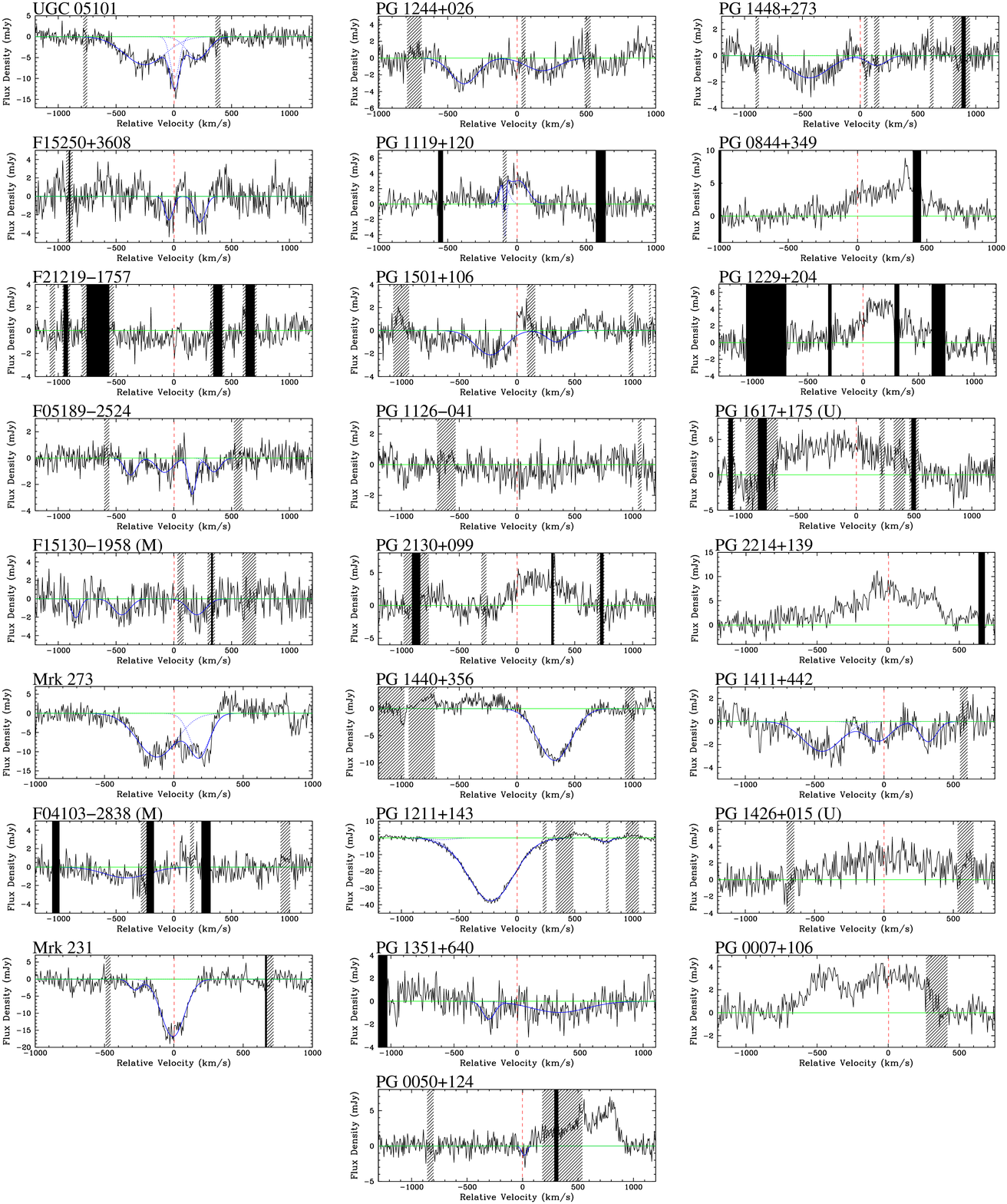}
\caption{Full intensity GBT spectra of 25 galaxies whose velocities did not overlap with RFI signals.  Columns from left to right are ordered by source type: ULIRGs -- FIR-strong quasars -- FIR-weak quasars.  Each column is organized with increasing infrared luminosity going down.  The targets with marginal or uncertain detections as noted in Section~\ref{sec:redux} are labeled with M and U, respectively, after their names.  The x-axis is displayed as velocity relative to the heliocentric radial velocity of the galaxy given in Table~\ref{tab:sample} and identified by the vertical dashed line.  The regions with hash marks are frequencies where the data are affected by persistent RFI spikes in some observing sessions, and the black regions are those where the data are affected by RFI spikes in all observing sessions.  The best Gaussian fits to the absorption features are highlighted in blue, where the dotted lines are individual Gaussian components (when there are multiple overlapping components) and the solid curves are the sums of these components.  
  The kinematics of these systems change from narrow absorption features likely due to outflows supported by the starbursts in ULIRGs, to broad absorption features from jet-driven outflows supported by the AGN in FIR-strong quasars, and finally to the emission profiles more characteristic of gas in rotation in the FIR-weak quasars. }
\label{fig:specs}
\end{figure}

\begin{figure}
\centering
\includegraphics[width=7in]{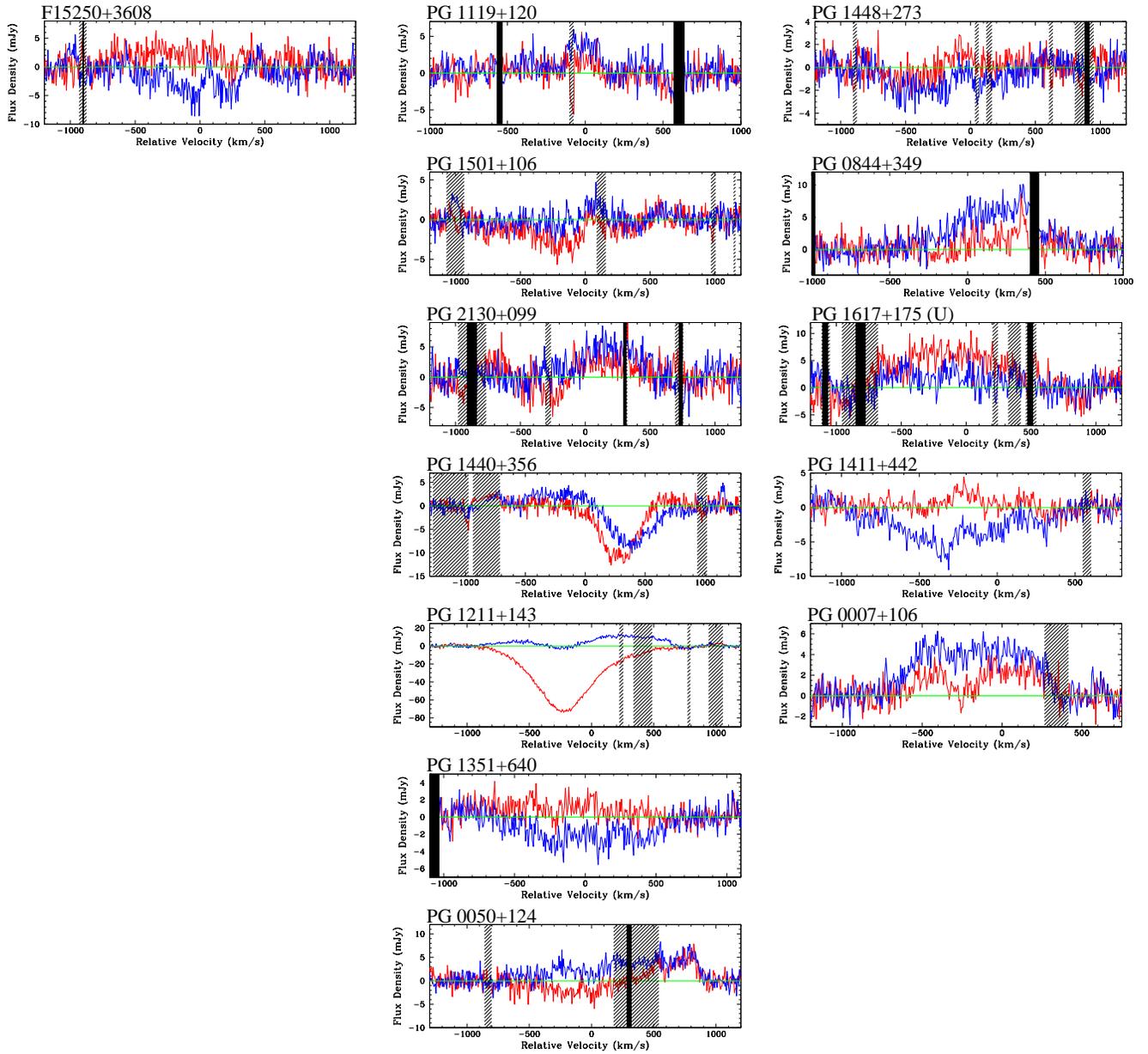}
\caption{The spectra of the 13 detected galaxies that show polarized HI spectra.  The organization of this figure is the same as that of Figure~\ref{fig:specs}.  The colors represent the two different polarizations (red: XX; blue: YY).  Polarized spectra are clearly seen in PG 0007+106, PG 0050+124, PG~1119+120, PG 1211+143, PG 1351+640, PG 1411+442, PG 1440+356,  and F15250+3608.  Weak polarized features may be present in PG 0844+349, PG 1448+273, PG 1501+106, PG 1617+175 (U), and PG 2130+099.  The FIR-strong quasars show the most distinct polarized features, perhaps reflecting the presence of polarized radio jets in these objects are contributing polarized continuum emission against which the absorption features are seen.}
\label{fig:polspecs}
\end{figure}

\begin{figure}
\centering
\includegraphics[width=7in]{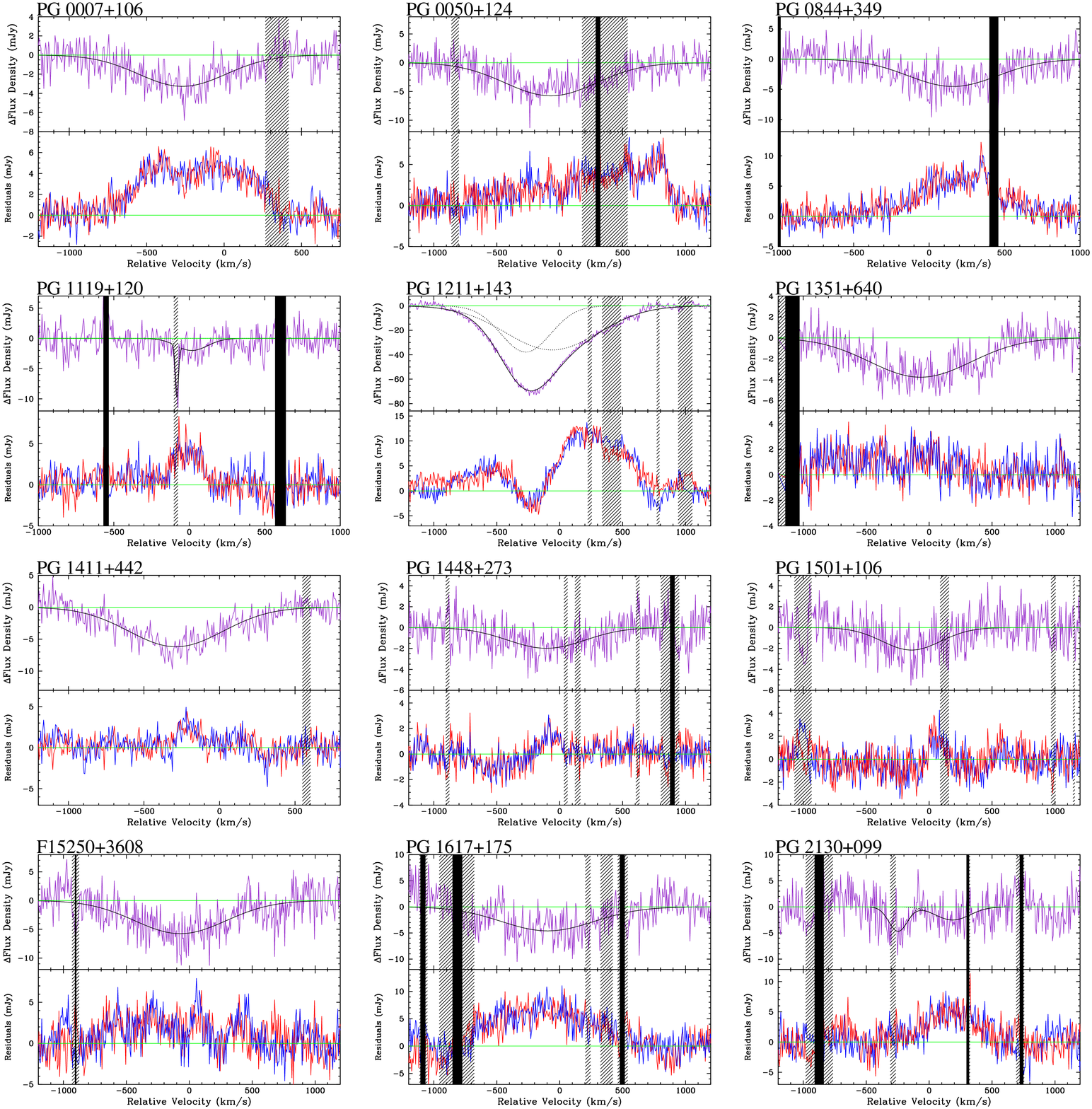}
\caption{Complex decomposition of the polarized HI spectra assuming that the difference in the two polarizations is due to more HI absorption in one polarization than the other.  For each galaxy, the upper panel is the difference spectrum between the two polarizations, {\em i.e.}, the spectrum in the polarization with the more negative flux density (Pol$_1$) minus the spectrum in the polarization with the higher flux density (Pol$_2$), plotted in purple.  The black curve shows the Gaussian model component(s) best fitting to the difference spectrum as tabulated in Table~\ref{tab:complex_absorption}.  The lower panel shows the residuals of the best fit model as applied to the Pol$_1$ spectrum and the Pol$_2$ spectrum to demonstrate the quality of the fit.  The colors in all of the bottom panels, the x-axis, and the hashed rectangles are the same as those in Figure~\ref{fig:polspecs}.  The polarization dependence of the HI spectra can be explained by broad HI absorption affecting one polarization more than the other.}
\label{fig:diffspecs}
\end{figure}

\begin{description}

\item[PG~0007+106]  The only radio-loud object in our sample based on classic radio-to-optical criteria.  This source is classified as a triple system in the third Zwicky catalog (III Zw 2), and the observed velocity range is consistent with the A ($v_{\rm sys} = 26921$~km~s$^{-1}$) and C ($v_{\rm sys} = 26981$~km~s$^{-1}$) components of the system.  The GBT detection is assumed to be that of the brightest component, A.  The HI emission profile of the full intensity spectrum agrees with the old Arecibo spectrum of \citet{hutchings87}.  The absorption feature, also noted by \citet{hutchings87}, is more prominent in the XX spectrum of the GBT data.  

\item[PG~0050+124 (I Zw 1)]  The heliocentric radial velocity listed in Table~\ref{tab:sample} for this source is from optical spectroscopy by \citet{ho09}.  The optically-derived velocity is $\sim$700~km~s$^{-1}$ lower than the HI-derived value of 18285 km~s$^{-1}$ \citep{bothun84}.  The optical emission lines may arise in part from outflowing gas, so the optically-derived value may is underestimate the actual systemic velocity of this object.  The shape of the full intensity profile (Figure~\ref{fig:specs}) at 18285~km~s$^{-1}$ agrees with those previously published by \citet{bothun84} and \citet{hutchings87}, and matches expectations for a regularly rotating disk.  However, at lower velocities, the profile becomes polarized and a broad absorption feature is centered at the optically-derived velocity (see Figure~\ref{fig:diffspecs}).  

\item[F04103-2838]  There is a marginal detection of an emission feature and a weak but broad blueshifted absorption feature present.  

\item[F05189-2524]  No emission is seen in the data.  However, several absorption components are detected.  An alternative explanation for the spectral shape is the superposition of an emission profile with a broad absorption component.

\item[PG~0804+761]  Strong RFI is present at the expected frequency of this galaxy.  It is therefore excluded from our analysis. 

\item[PG~0844+249]  The broad emission profile is asymmetric.  Although the profiles in the two different polarizations differ slightly, they have shapes similar to that observed by \citet{ho08}.  Our integrated GBT flux density is larger than that measured by \citet{ho08}.  The emission feature redward of the peak may be due to a companion at $v_{\rm sys} = 19601$~km~s$^{-1}$.  

\item[UGC~05101]  Multiple absorption components are seen in the GBT data.  We interpret the HI absorption profile as a highly absorbed core coincident with an inclined regularly rotating disk.  

\item[PG~1119+120]  This object was previously observed by \citet{bothun84}  and \citet{hutchings87} with Arecibo.  The profiles of the older spectra resemble our GBT observation, although the narrow absorption feature is not apparent in the lower-resolution Arecibo data.  There is no obvious detection of a companion in the GBT data near 16000~km~s$^{-1}$ as suggested by \citet{hutchings87}.  High resolution VLA maps at 4.8~GHz show a bent jet extending to the northeast  \citep{leipski}.  At the resolution of these VLA maps ($\sim$0.75$^{\prime \prime}$), the jet is resolved at a distance of $\sim$0.72~kpc.

\item[PG~1126-041]  This target was statistically undetected after three hours on-source, although there is a very weak broad absorption feature centered near its systemic velocity.

\item[PG~1211+143]  This broad absorption line system was undetected by \citet{condon85}; when it was observed by \citet{ho08} with Arecibo, standing waves were present in the data due to the strong continuum.

\item[PG~1229+204]  This object was observed by \citet{ho08}.  The profile of the GBT spectrum roughly agrees with the earlier data.

\item[PG~1244+026] A broad, blueshifted absorption feature is present.  As for F05189--2524, an alternative explanation for the spectral shape may be emission absorbed by a single broad absorption feature.  

\item[Mrk~231] The deep absorption trough is associated with the face-on disk of the galaxy \citep{carilli98, ulvestad99}.  Our GBT full intensity HI profile of this sources mostly agrees with these previous observations, but we detect an additional absorption feature centered at relative velocity $\simeq$--300~km~s$^{-1}$ that was first suspected by \citet{morganti11}.  This feature is slightly polarized, with the XX spectrum having lower intensity.  Although the polarization is not as high as those seen in Figure~\ref{fig:polspecs}, if we give the spectra of Mrk~231 the same treatment detailed in Section~\ref{sec:decompose}, we find that this feature has FWHM $\sim$1000~km~s$^{-1}$ corresponding to a mass outflow rate of $\sim$0.3~$M_\odot$~yr$^{-1}$ in the jet scenario.  The polarized feature is thus likely associated with the radio jet \citep{carilli98, ulvestad99}.  The velocity of this jet-enhanced outflow is similar to the value measured using Na~I in the jet-disturbed region \citep[--1400~km~s$^{-1}$;][]{rupke11}.  

\item[Mrk~273] This source is the only binary in our sample; its nuclear separation is $\sim$750~kpc \citep{scoville00}.  X-ray observations by \citet{iwasawa11} identified its southwestern nucleus as the Seyfert~2 nucleus and suggested that the northern nucleus may host a heavily obscured AGN.  The double-peaked HI absorption profile is very symmetric and is likely due to the two obscured nuclei.  There appears to be a weak absorption feature in the YY spectrum at --500 km~s$^{-1}$.  

\item[PG~1351+640]  The XX spectrum appears to show emission, while YY shows broad absorption.  The HI absorption data may also be variable.   \citet{leipski} note a jet-like structure in their 4.8 GHz VLA maps and time variability in the core flux density.  The resolution of the \citet{leipski} data ($\sim$0.56$^{\prime \prime}$) corresponds to a distance of $\sim$0.91~kpc.  Although on a different spatial scale, this structure is consistent (similar position angle) with a single-sided extension hinted at by the 8.4~GHz VLBA snapshot data presented by \citet{blundell}.  This feature is not seen in deeper VLBA imaging at multiple frequencies, but it should be noted that the data from \citet{ulvestad} are only in the left-circular polarization. 

\item[PG~1411+442]  The spectrum of this object is very similar to that of PG~1351+640.  Emission is seen in the XX spectrum and a two-component absorption feature in the YY spectrum.  

\item[PG~1426+015]  \citet{ho08} detect a fairly narrow ($W_{20}
= 357.4\,{\rm km\,s^{-1}}$) and well-defined HI line centered
at $v_{\rm sys} = 25975.2\,{\rm km\,s^{-1}}$.  We have examined our spectrum at the corresponding frequency and see hints of emission over a wider
range of velocities; however, this feature has such low significance
and is so sensitive to the details of baseline fitting we classify its
detection status as "uncertain."


\item[PG~1440+356]  The profile of the HI absorption feature depends slightly on the polarization.  Exceptionally, the feature is redshifted by $\sim$300~km~s$^{-1}$ with respect to systemic velocity rather than being blueshifted.  A possible explanation is that the systemic velocity taken from NED and based on optical emission lines underestimates the actual systemic velocity because the line-emitting gas arises in part from outflowing material.  Another possibility is that we may be detecting infalling HI.  Emission is also present in the blue half of the spectrum, although this is only seen at one of several epochs. 

\item[PG~1448+273]  The weakly polarized spectra show blueshifted absorption.  

\item[PG~1501+106]  A weak emission feature is present in the data.  A blueshifted absorption feature is also visible, particularly in the XX spectrum.  A lot of the data for this source were affected by small ripples in the baseline.  As a reality check, we compared our data with similar GBT data from program 10A-070 (PI: Kannappan).  The ripples also appear to be present in a large fraction of their data.  The persistent ripples could be due to harmonic waves from a strong RFI signal at a frequency that is slightly outside the detector frequency range.  Due to the expected weakness of the features, we conservatively flagged all of the data that showed these ripples.

\item[F15130-1958]  Multiple weak absorption components are present in this object.

\item[F15250+3608]  This is the only ULIRG in our sample that shows distinct polarization-dependent HI absorption.  The spectra are very similar to those of PG~1351+640 and PG~1411+442, where weak emission is seen in the XX spectrum and absorption in YY.  

\item[F15462-0450]  Strong RFI is present at the expected observed frequency of this galaxy.  It is therefore excluded from our analysis.

\item[PG~1617+175]  Broad emission is detected in this object.  The profiles of the emission differ slightly in the two polarizations. Notably, the emission feature in the YY spectrum seems to be affected by an additional absorption feature (Figure~\ref{fig:diffspecs}).  

\item[F21219-1757]  This object was not detected.

\item[PG~2130+099] This target was observed by \citet{ho08}.  Our full intensity profile of this source agrees with the older data from Arecibo.  A blueshifted absorption feature is also seen in the XX spectrum.  \citet{leipski} detected what looks like jet-associated hot spots emanating from the core in 4.8~GHz A-array VLA data.  The resolution of these VLA data ($\sim$0.35$^{\prime \prime}$) implies that the jet can be resolved at a distance of $\sim$0.42~kpc.

\item[PG~2214+139]  The HI spectrum of this object is dominated by a broad emission feature centered at systemic velocity.

\end{description}


\begin{turnpage}
\begin{deluxetable}{lccccccrccccccc}
\tablecolumns{15}
\tabletypesize{\scriptsize}
\tablecaption{The Sample and Journal of Observations}
\tablewidth{0.0pt}
\tablehead{\colhead{Source} & \colhead{RA} &\colhead{Dec} & \colhead{$v_{\rm sys}$}  & \colhead{Type} & \colhead{FIR} & \colhead{$\log (\frac{L_{\rm bol}}{L_\odot})$} &\colhead{SFR}& \colhead{$S_{\rm 1.4~GHz}$} & \colhead{$\log R_X$} & \colhead{Dates of} & \colhead{t$_{\rm int}$} &\colhead{$\Delta V$} &\colhead{$\sigma_{\rm chan}$}& \colhead{Fit}\\
\colhead{} & \colhead{J2000} & \colhead{J2000} &\colhead{(km s$^{-1}$)}  &\colhead{}  & \colhead{Strength} & \colhead{} &\colhead{($M_\odot$~yr$^{-1}$)}& \colhead{(Jy)} & \colhead{} & \colhead{Observations} & \colhead{(s)} &\colhead{(km~s$^{-1}$)}&\colhead{(mJy)}& \colhead{Order}\\
\colhead{(1)} & \colhead{(2)} & \colhead{(3)} & \colhead{(4)} & \colhead{(5)} & \colhead{(6)} & \colhead{(7)} & \colhead{(8)} & \colhead{(9)} & \colhead{(10)} & \colhead{(11)} & \colhead{(12)}&\colhead{(13)}&\colhead{(14)}&\colhead{(15)} 
}
\startdata
PG 0007+106	&	00:10:31.0	&	+10:58:30	&	26783	&	QSO	&	Weak	&	12.23	&	$\sim$1.2	&	0.099$^1$	&	--2.4	&	2011 Sept -- 2012 Jan	&	7894	&	6.11	&	0.77	&	3	\\
PG 0050+124	&	00:53:34.9	&	+12:41:36	&	17658	&	QSO	&	Strong	&	12.07	&	10.4	&	0.009$^1$	&	--5.0	&	2011 Sept -- 2012 Jan	&	4815	&	5.78	&	0.94	&	3	\\
F04103--2838	&	04:12:19.4	&	--28:30:25	&	35215	&	ULIRG	&	Strong	&	12.30	&	154.2	&	0.013$^1$	&	--3.4	&	2011 Sept -- 2012 Mar	&	7452	&	6.43	&	0.88	&	3	\\
F05189--2524	&	05:21:01.4	&	--25:21:45	&	12760	&	ULIRG	&	Strong	&	12.22	&	71.6	&	0.029$^1$	&	\nodata	&	2011 Sept -- 2012 Jan	&	12330	&	5.60	&	0.58	&	3	\\
PG 0804+761	&	08:10:58.6	&	+76:02:43	&	29979	&	QSO	&	Strong	&	12.08	&	0.3	&	0.004$^1$	&	--5.3	&	2011 Aug	&	860	&	6.23	&	2.97	&	\nodata	\\
PG 0844+349	&	08:47:42.4	&	+34:45:04	&	19187	&	QSO	&	Weak	&	11.44	&	0.8	&	$<$ 0.0025$^1$	&	$<$--3.3	&	2011 Nov	&	3153	&	5.83	&	0.96	&	3	\\
UGC 05101	&	09:35:51.6	&	+61:21:11	&	11802	&	ULIRG	&	Strong	&	12.05	&	110.2	&	0.170	&	--3.0	&	2011 Aug -- 2011 Sept	&	2151	&	5.57	&	1.46	&	3	\\
PG 1119+120	&	11:21:47.1	&	+11:44:18	&	15050	&	QSO	&	Strong	&	11.33	&	0.4	&	0.004$^1$	&	\nodata	&	2011 Oct -- 2012 Mar	&	3837	&	5.68	&	0.83	&	3	\\
PG 1126--041	&	11:29:16.6	&	--04:24:08	&	18575	&	QSO	&	Strong	&	11.52	&	2.5	&	$<$ 0.0025$^1$	&	$<$--4.6	&	2011 Aug -- 2012 Jan	&	8871	&	5.81	&	0.54	&	3	\\
PG 1211+143	&	12:14:17.7	&	+14:03:13	&	24253	&	QSO	&	Strong	&	11.96	&	0.0	&	0.600	&	--2.8	&	2011 Aug -- 2012 Mar	&	5986	&	6.02	&	0.92	&	3	\\
PG 1229+204	&	12:32:03.6	&	+20:09:29	&	18890	&	QSO	&	Weak	&	11.56	&	0.4	&	0.003$^1$	&	--4.9	&	2012 Mar	&	3411	&	5.82	&	0.92	&	3	\\
PG 1244+026	&	12:46:35.2	&	+02:22:09	&	14443	&	QSO	&	Strong	&	11.02	&	1.5	&	0.002$^2$	&	--5.1	&	2011 Aug -- 2012 Mar	&	3555	&	5.66	&	1.00	&	3	\\
Mrk 231	&	12:56:14.2	&	+56:52:25	&	12642	&	ULIRG	&	Strong	&	12.60	&	174.2	&	0.309	&	--2.7	&	2011 Aug -- 2011 Sept	&	1257	&	5.60	&	1.52	&	3	\\
Mrk 273	&	13:44:42.1	&	+55:53:13	&	11326	&	ULIRG	&	Strong	&	12.24	&	141.6	&	0.136	&	--3.5	&	2011 Aug -- 2011 Sept	&	1290	&	5.50	&	1.31	&	3	\\
PG 1351+640	&	13:53:15.8	&	+63:45:46	&	26442	&	QSO	&	Strong	&	12.04	&	26.1	&	0.034$^1$	&	--3.3	&	2011 Sept -- 2012 Feb	&	8153	&	6.10	&	0.79	&	3	\\
PG 1411+442	&	14:13:48.3	&	+44:00:14	&	26861	&	QSO	&	Weak	&	11.78	&	0.0	&	$<$ 0.0025$^1$	&	$<$--4.6	&	2011 Aug -- 2012 Feb	&	6346	&	6.12	&	0.92	&	3	\\
PG 1426+015	&	14:29:06.6	&	+01:17:06	&	25953	&	QSO	&	Weak	&	11.92	&	1.9	&	0.004$^1$	&	--5.0	&	2012 Mar	&	3153	&	6.08	&	1.09	&	5	\\
PG 1440+356	&	14:42:07.4	&	+35:26:23	&	23700	&	QSO	&	Strong	&	11.80	&	10.3	&	0.005$^1$	&	--4.8	&	2011 Aug -- 2011 Dec	&	6135	&	6.00	&	0.63	&	3	\\
PG 1448+273	&	14:51:08.7	&	+27:09:27	&	19487	&	QSO	&	Weak	&	11.43	&	0.1	&	0.003$^3$	&--4.8	&	2011 Aug -- 2011 Feb	&	8917	&	5.85	&	0.63	&	3	\\
PG 1501+106	&	15:04:01.2	&	+10:26:16	&	10919	&	QSO	&	Strong	&	11.33	&	0.0	&	$<$ 0.0025$^1$	&	$<$--5.7	&	2012 Feb -- 2012 Mar	&	6540	&	5.53	&	0.77	&	3	\\
F15130--1958	&	15:15:55.2	&	--20:09:17	&	32585	&	ULIRG	&	Strong	&	12.23	&	108.8	&	0.010$^1$	&	--2.4	&	2011 Dec -- 2012 Mar	&	2666	&	6.33	&	1.25	&	3	\\
F15250+3608	&	15:26:59.4	&	+35:58:38	&	16535	&	ULIRG	&	Strong	&	12.12	&	109.6	&	0.018	&	\nodata	&	2011 Aug -- 2011 Nov	&	2007	&	5.74	&	1.16	&	3	\\
F15462--0450	&	15:28:56.8	&	--04:59:34	&	29917	&	ULIRG	&	Strong	&	12.28	&	168.7	&	0.013$^1$	&	--3.4	&	2011 Dec	&	860	&	6.23	&	4.14	&	\nodata	\\
PG 1617+175	&	16:20:11.3	&	+17:24:28	&	33708	&	QSO	&	Weak	&	11.74	&	4.0	&	$<$ 0.0025$^1$	&	\nodata	&	2011 Aug -- 2012 Mar	&	2547	&	6.38	&	1.59	&	3	\\
F21219--1757	&	21:24:41.6	&	--17:44:46	&	33526	&	ULIRG	&	Strong	&	12.17	&	48.3	&	0.015$^1$	&	\nodata	&	2011 Oct -- 2012 Mar	&	8529	&	6.37	&	0.70	&	3	\\
PG 2130+099	&	21:32:27.8	&	+10:08:19	&	18880	&	QSO	&	Strong	&	11.77	&	0.2	&	0.007$^1$	&	--2.7	&	2012 Mar	&	2001	&	5.82	&	1.13	&	5	\\
PG 2214+139	&	22:17:12.2	&	+14:14:21	&	19715	&	QSO	&	Weak	&	11.77	&	0.0	&	$<$ 0.0025$^1$	&	$<$--5.0	&	2011 Sept -- 2011 Oct	&	2121	&	5.85	&	1.32	&	3	\\
\enddata
\tablerefs{$^1$\citet[NVSS;][]{nvss}; $^2$\citet{rafter}; $^3$\citet{barvainis}.}
\tablecomments{
Col.(1): Galaxy name.  Coordinate-based names beginning with ``F'' are sources in the {\it IRAS} Faint Source Catalog; the ``PG'' sources are Palomar Green quasars.  Col.(2): Right ascension in J2000 coordinates.  Col. (3): Declination in J2000 coordinates.  Col.(4): Heliocentric radial velocity given by NED.  These velocities are measured via optical emission lines and may underestimate the true recessional velocities since some fraction of the line emission may arise from the outflowing material itself.  All velocities are quoted in the optical convention where $v=cz$.  Col.(5): ULIRG or QSO source.  Col.(6):  FIR strength as defined by \citet{netzer07} where FIR-strong sources have ${f_{\rm 60 \mu m}}/{f_{\rm 15 \mu m}} > 1$.   Col.(7): Bolometric luminosity.  For ULIRGs, we assume $L_{\rm bol} = 1.15 L_{\rm IR}$.  For PG QSOs, we assume $L_{\rm bol} = 7 L(5100~\AA)$ \citep{netzer07}.  Col.(8):  Star formation rates in solar masses per year derived using the average AGN contribution fraction to the total bolometric luminosity measured by \citet{vei09a} for each galaxy.  For PG~0007+106, the AGN fraction is assumed to be the mean for all other FIR-weak quasars.  It is assumed that all of the emission from the starburst is reprocessed by dust and re-emitted into the IR band.  The SFR is derived using the relation given in \citet{ken98} for a 0.1--100~$M_\odot$ initial mass function.  Col.(9):  1.4~GHz continuum flux densities.  Values taken with the GBT unless noted.  The upper limits are given as the detection limit of the NRAO VLA Sky Survey (NVSS).  Col.(10):  1.4~GHz to 2--10~keV flux ratio, where $\log R_X > -4.3$, implies a radio-loud source \citep{lafranca}.  We used the (average, if multiple observations were performed) absorption-corrected 2--10~keV flux from \citet{teng10} except for PG~0007+106 and PG~0804+761 whose X-ray measurements were taken from \citet{p05}.   These ratios are corrected for the 1.4~GHz contribution from star formation based on the SFRs listed in Table~\ref{tab:sample} and the SFR-1.4~GHz relation from \citet{bell03}, which assumes the same 0.1--100~$M_\odot$ initial mass function as \citet{ken98}.  For ULIRGs, due to their high line-of-sight column densities, their measured 2--10~keV fluxes may be underestimated and thus the listed values may represent upper limits.   Col.(11):  Period over which the target was observed.  Col.(12):  Total on source time in seconds after flagging.  Col.(13):  Velocity resolution of the full intensity spectrum.  Col.(14):  The channel-to-channel rms of the full intensity spectrum.  Col.(15):  The order of polynomial over which the baseline was fitted and then subtracted.
}
\label{tab:sample}
\end{deluxetable}
\end{turnpage}

\begin{deluxetable}{lcccrccccrc}
\tablecolumns{9}
\tabletypesize{\scriptsize}
\tablecaption{Measurements of HI in Emission}
\tablewidth{0pt}
\tablehead{\colhead{Source} & \colhead{$V_\odot$} & \colhead{$W_{20}$}  & \colhead{$W_{50}$} & \colhead{Peak} &\colhead{S/N}& \colhead{$f_{\rm HI}$} &\colhead{$N_{\rm HI}$}&\colhead{$M_{\rm HI}$}\\
\colhead{} & \colhead{(km~s$^{-1}$)} & \colhead{(km~s$^{-1}$)} &\colhead{(km s$^{-1}$)}  &\colhead{(mJy)}  &\colhead{}& \colhead{(Jy km~s$^{-1}$)} &\colhead{(10$^{18}$~cm$^{-2}$)}&\colhead{(10$^{10}~M_\odot$)}\\
\colhead{(1)} & \colhead{(2)} & \colhead{(3)} & \colhead{(4)} & \colhead{(5)} & \colhead{(6)} & \colhead{(7)} & \colhead{(8)} & \colhead{(9)} 
}
\startdata
PG 0007+106	& $	-216	\pm	16	$ & $	1063	\pm	49	$ & $	894	\pm	33	$ &	4.30	&	5.58	& $	2.57	\pm	0.14	$ & $	9.35	\pm	0.49	$ & $	8.46	\pm	0.45	$ \\
PG 0050+123	& $	473	\pm	17	$ & $	789	\pm	51	$ & $	435	\pm	34	$ &	7.00	&	7.45	& $	2.43	\pm	0.14	$ & $	8.85	\pm	0.51	$ & $	3.44	\pm	0.20	$ \\
F04103-2838 (M)	& $	129	\pm	21	$ & $	151	\pm	63	$ & $	21	\pm	42	$ &	3.44	&	3.91	& $	0.17	\pm	0.06	$ & $	0.62	\pm	0.22	$ & $	0.97	\pm	0.34	$ \\
PG 0844+249	& $	243	\pm	8	$ & $	541	\pm	23	$ & $	436	\pm	15	$ &	8.83	&	9.20	& $	2.32	\pm	0.12	$ & $	8.44	\pm	0.43	$ & $	4.05	\pm	0.21	$ \\
PG 1119+120	& $	-18	\pm	2	$ & $	226	\pm	7	$ & $	221	\pm	5	$ &	5.35	&	6.45	& $	0.58	\pm	0.07	$ & $	2.11	\pm	0.24	$ & $	0.65	\pm	0.07	$ \\
PG 1211+143	& $	522	\pm	8	$ & $	200	\pm	25	$ & $	174	\pm	17	$ &	3.85	&	4.18	& $	0.04	\pm	0.07	$ & $	0.14	\pm	0.25	$ & $	0.11	\pm	0.20	$ \\
PG 1229+204	& $	182	\pm	11	$ & $	600	\pm	34	$ & $	504	\pm	23	$ &	5.37	&	5.84	& $	1.84	\pm	0.12	$ & $	6.70	\pm	0.43	$ & $	3.24	\pm	0.21	$ \\
Mrk 273	& $	407	\pm	11	$ & $	116	\pm	32	$ & $	62	\pm	21	$ &	5.94	&	4.53	& $	0.37	\pm	0.07	$ & $	1.35	\pm	0.26	$ & $	0.24	\pm	0.05	$ \\
PG 1351+640	& $	-205	\pm	21	$ & $	585	\pm	64	$ & $	480	\pm	43	$ &	2.64	&	3.34	& $	0.82	\pm	0.10	$ & $	2.98	\pm	0.38	$ & $	2.73	\pm	0.34	$ \\
PG 1426+015 (U)	& $	107	\pm	35	$ & $	1131	\pm	104	$ & $	591	\pm	69	$ &	5.12	&	4.69	& $	2.08	\pm	0.20	$ & $	7.57	\pm	0.72	$ & $	6.86	\pm	0.65	$ \\
PG 1440+356	& $	-294	\pm	13	$ & $	477	\pm	39	$ & $	395	\pm	26	$ &	3.01	&	4.81	& $	0.48	\pm	0.07	$ & $	1.73	\pm	0.27	$ & $	1.30	\pm	0.20	$ \\
PG 1448+273	& $	-50	\pm	18	$ & $	97	\pm	54	$ & $	23	\pm	36	$ &	2.04	&	3.26	& $	0.11	\pm	0.03	$ & $	0.41	\pm	0.12	$ & $	0.21	\pm	0.06	$ \\
PG 1501+106	& $	61	\pm	21	$ & $	137	\pm	63	$ & $	22	\pm	42	$ &	2.58	&	3.37	& $	0.32	\pm	0.05	$ & $	1.16	\pm	0.17	$ & $	0.20	\pm	0.03	$ \\
PG 1617+175 (U)	& $	-118	\pm	15	$ & $	1139	\pm	44	$ & $	1057	\pm	30	$ &	6.94	&	4.36	& $	3.88	\pm	0.30	$ & $	14.12	\pm	1.08	$ & $	21.21	\pm	1.62	$ \\
PG 2130+099	& $	224	\pm	9	$ & $	599	\pm	26	$ & $	518	\pm	18	$ &	7.86	&	6.95	& $	2.30	\pm	0.15	$ & $	8.37	\pm	0.53	$ & $	3.84	\pm	0.24	$ \\
PG 2214+139	& $	104	\pm	14	$ & $	876	\pm	42	$ & $	567	\pm	28	$ &	11.22	&	8.53	& $	4.43	\pm	0.21	$ & $	16.13	\pm	0.75	$ & $	7.99	\pm	0.37	$ \\
\enddata
\tablecomments{
Col.(1): Galaxy name where (M) and (U) indicate marginal and uncertain detections, respectively, as noted in Section~\ref{sec:redux}.   Col.(2):  Heliocentric velocity measured from HI relative to the systematic velocity presented in Table~\ref{tab:sample}, and its associated measurement error.  Col.(3):  Velocity width of the emission measured at the 20\% level and its error.  Col.(4):  Velocity width of the emission measured at the 50\% level  and its error.  Col.(5):  Maximum intensity of the spectrum.  Col.(6):  Signal-to-noise ratio given by the peak intensity relative to the channel rms listed in Table~\ref{tab:sample}.  Col.(7):  Velocity-integrated HI flux in emission and its error.  Col.(8):  Derived HI column density calculated using Equation~1.  Col.(9):  Derived HI gas mass from emission and its error calculated using Equation~2.
}
\label{tab:emission}
\end{deluxetable}

\begin{deluxetable}{lcccrrcc}
\tablecolumns{8}
\tablecaption{Measurements of HI in Absorption}
\tablewidth{0pt}
\tablehead{\colhead{Source} & \colhead{$V_\odot$} & \colhead{FWHM}  & \colhead{``Peak''}  &\colhead{S/N} &\colhead{$N_{\rm HI}$} & \colhead{$\tau_{\rm max}$} & \colhead{$f_{\rm abs}$}\\
\colhead{} &\colhead{(km~s$^{-1}$)}  &\colhead{(km~s$^{-1}$)} & \colhead{(mJy)} &\colhead{}& \colhead{(10$^{22}$~cm~$^{-2}$)} & \colhead{} & \colhead{(Jy km~s$^{-1}$)}\\
\colhead{(1)} & \colhead{(2)} & \colhead{(3)} & \colhead{(4)} & \colhead{(5)} & \colhead{(6)} & \colhead{(7)} & \colhead{(8)} 
}
\startdata
\multicolumn{8}{c}{Strong Detections: peak-to-rms $> 3 \sigma$}\\ \hline
F05189--2524	& $	160	\pm	3	$ & $	79	\pm	8	$ & $	-2.74	\pm	0.23	$ &	4.73	&	1.50	&	0.099	&	--0.04	\\
UGC 05101	& $	-251	\pm	11	$ & $	419	\pm	26	$ & $	-6.69	\pm	0.22	$ &	4.59	&	3.24	&	0.040	&	--0.54	\\
	& $	8	\pm	3	$ & $	96	\pm	7	$ & $	-9.76	\pm	0.59	$ &	6.70	&	1.09	&	0.059	&	--0.18	\\
	& $	203	\pm	9	$ & $	219	\pm	22	$ & $	-5.23	\pm	0.34	$ &	3.59	&	1.32	&	0.031	&	--0.22	\\
PG 1119+120	& $	-88	\pm	3	$ & $	22	\pm	6	$ & $	-4.20	\pm	0.87	$ &	5.06	&	\nodata	&	\nodata	&	--0.02	\\
PG 1211+143	& $	-227	\pm	1	$ & $	460	\pm	3	$ & $	-37.67	\pm	0.16	$ &	40.95	&	5.73	&	0.065	&	--3.07	\\
PG 1244+026	& $	-381	\pm	8	$ & $	226	\pm	20	$ & $	-3.07	\pm	0.23	$ &	3.05	&	\nodata	&	\nodata	&	--0.13	\\
Mrk 231	& $	-6	\pm	2	$ & $	199	\pm	6	$ & $	-16.87	\pm	0.41	$ &	11.07	&	2.15	&	0.056	&	--0.64	\\
Mrk 273	& $	-122	\pm	8	$ & $	334	\pm	21	$ & $	-11.41	\pm	0.41	$ &	8.71	&	5.60	&	0.088	&	--0.74	\\
	& $	187	\pm	6	$ & $	171	\pm	13	$ & $	-10.62	\pm	0.64	$ &	8.11	&	2.66	&	0.081	&	--0.35	\\
PG 1440+356	& $	325	\pm	4	$ & $	333	\pm	8	$ & $	-9.57	\pm	0.21	$ &	15.27	&	\nodata	&	\nodata	&	--0.57	\\
\cutinhead{Weak Detections: peak-to-rms $< 3\sigma$}\\ 	
PG 0050+124	& $	15	\pm	10	$ & $	53	\pm	24	$ & $	-1.34	\pm	0.51	$ &	1.42	&	1.62	&	0.161	&	--0.01	\\
F04103--2838 (M)	& $	-409	\pm	29	$ & $	449	\pm	70	$ & $	-1.17	\pm	0.16	$ &	1.33	&	8.10	&	0.094	&	--0.09	\\
F05189--2524	& $	-374	\pm	9	$ & $	118	\pm	22	$ & $	-1.29	\pm	0.20	$ &	2.22	&	1.03	&	0.046	&	--0.03	\\
	& $	-79	\pm	13	$ & $	163	\pm	30	$ & $	-1.09	\pm	0.17	$ &	1.88	&	1.20	&	0.038	&	--0.03	\\
	& $	347	\pm	11	$ & $	127	\pm	27	$ & $	-1.05	\pm	0.19	$ &	1.81	&	0.90	&	0.037	&	--0.03	\\
PG 1211+143	& $	-644	\pm	25	$ & $	214	\pm	60	$ & $	-1.14	\pm	0.25	$ &	1.24	&	0.08	&	0.002	&	--0.04	\\
	& $	766	\pm	11	$ & $	116	\pm	26	$ & $	-2.47	\pm	0.48	$ &	2.68	&	0.09	&	0.004	&	--0.05	\\
PG 1244+026	& $	183	\pm	18	$ & $	268	\pm	43	$ & $	-1.52	\pm	0.21	$ &	1.52	&	59.07	&	1.407	&	--0.08	\\
Mrk 231	& $	-280	\pm	9	$ & $	96	\pm	21	$ & $	-3.14	\pm	0.58	$ &	2.06	&	0.19	&	0.010	&	--0.06	\\
PG 1351+640	& $	-224	\pm	10	$ & $	115	\pm	25	$ & $	-1.53	\pm	0.28	$ &	1.94	&	1.02	&	0.046	&	--0.03	\\
	& $	336	\pm	37	$ & $	516	\pm	98	$ & $	-0.97	\pm	0.14	$ &	1.23	&	2.88	&	0.029	&	--0.09	\\
PG 1411+442	& $	-446	\pm	10	$ & $	319	\pm	23	$ & $	-2.60	\pm	0.16	$ &	2.81	&	\nodata	&	\nodata	&	--0.14	\\
	& $	-41	\pm	10	$ & $	214	\pm	25	$ & $	-1.70	\pm	0.17	$ &	1.84	&	$>$39.78	&	1.139	&	--0.06	\\
	& $	319	\pm	10	$ & $	143	\pm	24	$ & $	-1.76	\pm	0.26	$ &	1.90	&	$>$28.08	&	1.217	&	--0.04	\\
PG 1448+273	& $	-441	\pm	14	$ & $	392	\pm	34	$ & $	-1.69	\pm	0.13	$ &	2.70	&	55.58	&	0.829	&	--0.12	\\
	& $	144	\pm	19	$ & $	195	\pm	45	$ & $	-0.74	\pm	0.15	$ &	1.18	&	10.27	&	0.283	&	--0.03	\\
PG 1501+106	& $	-229	\pm	12	$ & $	304	\pm	28	$ & $	-2.12	\pm	0.17	$ &	2.77	&	$>$83.12	&	1.884	&	--0.12	\\
	& $	340	\pm	20	$ & $	195	\pm	48	$ & $	-0.98	\pm	0.21	$ &	1.28	&	$>$17.46	&	0.498	&	--0.04	\\
F15130--1958 (M)	& $	-847	\pm	10	$ & $	89	\pm	23	$ & $	-2.07	\pm	0.47	$ &	1.66	&	3.86	&	0.232	&	--0.03	\\
	& $	-451	\pm	15	$ & $	168	\pm	36	$ & $	-1.75	\pm	0.32	$ &	1.40	&	6.10	&	0.192	&	--0.05	\\
	& $	199	\pm	16	$ & $	182	\pm	38	$ & $	-1.74	\pm	0.32	$ &	1.40	&	6.56	&	0.191	&	--0.05	\\
F15250+3608	& $	-42	\pm	8	$ & $	83	\pm	18	$ & $	-2.57	\pm	0.49	$ &	2.21	&	2.42	&	0.154	&	--0.04	\\
	& $	229	\pm	8	$ & $	107	\pm	19	$ & $	-2.79	\pm	0.43	$ &	2.40	&	3.41	&	0.168	&	--0.06	\\
\enddata
\tablecomments{
Col.(1): Galaxy name.  Col.(2):  Heliocentric velocity measured from HI relative to the systematic velocity presented in Table~\ref{tab:sample}, as modeled using a Gaussian model, and its associated measurement error.  Col.(3):  Full width of the absorption feature at the 50\% flux level as measured using a Gaussian model and its error.  Col.(4):  Peak absorption flux and its error as measured using a Gaussian model.  Col.(5):  Signal-to-noise ratio measured as a ratio between the peak flux and channel rms listed in Table~\ref{tab:sample}.  Col.(6):  Integrated HI column density based on the absorption model, assuming a spin temperature of 1000~K and using Equation 4 based on the optical depth profile derived using Equation 3.  For PG~1411+442 and PG~1501+106, the 1.4~GHz core flux densities are only upper limits, so their derived $N_{\rm HI}$ values are lower limits.  Col.(7):  Peak optical depth based on Equation~3.  Col.(8):  Velocity-integrated HI flux in absorption measured from the best-fit Gaussian component models. 
}
\label{tab:absorption}
\end{deluxetable}

\begin{deluxetable}{lrrrrccccrc}
\tablecolumns{9}
\tabletypesize{\scriptsize}
\tablecaption{Measurements of HI in Emission from Complex Spectral Modeling}
\tablewidth{0pt}
\tablehead{\colhead{Source} & \colhead{$V_\odot$} & \colhead{$W_{20}$}  & \colhead{$W_{50}$}  & \colhead{``Peak''} &\colhead{S/N}& \colhead{$f_{\rm HI}$} &\colhead{$N_{\rm HI}$}&\colhead{$M_{\rm HI}$}\\
\colhead{} &\colhead{(km~s$^{-1}$)}  & \colhead{(km~s$^{-1}$)} & \colhead{(km~s$^{-1}$)} & \colhead{(mJy)} &\colhead{}& \colhead{(Jy km~s$^{-1}$)} &\colhead{(10$^{18}$~cm$^{-2}$)}&\colhead{(10$^{10}~M_\odot$)}\\
\colhead{(1)} & \colhead{(2)} & \colhead{(3)} & \colhead{(4)} & \colhead{(5)} & \colhead{(6)} & \colhead{(7)} & \colhead{(8)} & \colhead{(9)} 
}
\startdata
PG 0007+106	& $	-211	\pm	11	$ & $	1054	\pm	33	$ & $	917	\pm	22	$ &	5.74	&	7.41	& $	3.81	\pm	0.14	$ & $	13.86	\pm	0.50	$ & $	12.53	\pm	0.45	$ \\
PG 0050+123	& $	184	\pm	21	$ & $	1367	\pm	64	$ & $	916	\pm	42	$ &	8.32	&	6.80	& $	4.53	\pm	0.24	$ & $	16.48	\pm	0.87	$ & $	6.40	\pm	0.34	$ \\
PG 0844+249	& $	120	\pm	14	$ & $	921	\pm	43	$ & $	474	\pm	29	$ &	10.65	&	10.06	& $	4.36	\pm	0.17	$ & $	15.86	\pm	0.62	$ & $	7.61	\pm	0.30	$ \\
PG 1119+120	& $	-15	\pm	3	$ & $	232	\pm	8	$ & $	222	\pm	6	$ &	6.23	&	7.43	& $	0.90	\pm	0.07	$ & $	3.28	\pm	0.24	$ & $	1.02	\pm	0.08	$ \\
PG 1211+143	& $	-235	\pm	17	$ & $	1802	\pm	50	$ & $	655	\pm	34	$ &	13.10	&	13.97	& $	7.89	\pm	0.21	$ & $	28.73	\pm	0.78	$ & $	22.61	\pm	0.67	$ \\
PG 1351+640	& $	-332	\pm	29	$ & $	1202	\pm	88	$ & $	807	\pm	59	$ &	3.56	&	4.72	& $	1.34	\pm	0.14	$ & $	4.89	\pm	0.52	$ & $	4.48	\pm	0.47	$ \\
PG 1411+442	& $	-205	\pm	14	$ & $	208	\pm	41	$ & $	169	\pm	27	$ &	3.22	&	3.19	& $	0.42	\pm	0.08	$ & $	1.52	\pm	0.29	$ & $	1.45	\pm	0.28	$ \\
PG 1448+273	& $	-68	\pm	12	$ & $	168	\pm	37	$ & $	91	\pm	25	$ &	3.07	&	4.83	& $	0.26	\pm	0.04	$ & $	0.93	\pm	0.16	$ & $	0.48	\pm	0.08	$ \\
PG 1501+106	& $	61	\pm	5	$ & $	135	\pm	16	$ & $	122	\pm	11	$ &	3.22	&	4.53	& $	0.22	\pm	0.04	$ & $	0.81	\pm	0.16	$ & $	0.14	\pm	0.03	$ \\
F15250+3608	& $	-89	\pm	10	$ & $	1151	\pm	30	$ & $	1122	\pm	20	$ &	5.44	&	3.67	& $	2.38	\pm	0.26	$ & $	8.66	\pm	0.96	$ & $	3.23	\pm	0.36	$ \\
PG 1617+175 (U)	& $	-88	\pm	13	$ & $	1315	\pm	38	$ & $	1208	\pm	25	$ &	9.23	&	5.81	& $	6.10	\pm	0.32	$ & $	22.19	\pm	1.16	$ & $	33.32	\pm	1.75	$ \\
PG 2130+099	& $	67	\pm	20	$ & $	909	\pm	61	$ & $	366	\pm	40	$ &	8.56	&	7.87	& $	2.62	\pm	0.17	$ & $	9.54	\pm	0.63	$ & $	4.37	\pm	0.29	$ \\
\enddata
\tablecomments{
Col.(1): Galaxy name.  Col.(2):  Heliocentric velocity measured from HI relative to the systematic velocity presented in Table~\ref{tab:sample}, and its associated measurement error.  Col.(3):  Velocity width of the emission measured at the 20\% level and its error.  Col.(4):  Velocity width of the emission measured at the 50\% level  and its error.  Col.(5):  The maximum intensity of the spectrum.  Col.(6):  Signal-to-noise ratio given by the peak intensity relative to the channel rms listed in Table~\ref{tab:sample}.  Col.(7):  Velocity-integrated HI flux in emission and its error.  Col.(8):  Derived HI column density calculated using Equation~1.  Col.(9):  Derived HI gas mass from emission and its error, calculated using Equation~2.
}
\label{tab:complex_emission}
\end{deluxetable}

\begin{deluxetable}{lccrrrcc}
\tablecolumns{8}
\tablecaption{Measurements of HI in Absorption from Complex Spectral Modeling}
\tablewidth{0pt}
\tablehead{\colhead{Source} & \colhead{$V_\odot$} & \colhead{FWHM}  & \colhead{``Peak''} &\colhead{S/N} &\colhead{$N_{\rm HI}$} & \colhead{$\tau_{\rm max}$} & \colhead{$f_{\rm abs}$}\\
\colhead{} & \colhead{(km~s$^{-1}$)} &\colhead{(km s$^{-1}$)} & \colhead{(mJy)} &\colhead{}& \colhead{(10$^{22}$~cm~$^{-2}$)} & \colhead{} & \colhead{(Jy km~s$^{-1}$)}\\
\colhead{(1)} & \colhead{(2)} & \colhead{(3)} & \colhead{(4)} & \colhead{(5)} & \colhead{(6)} & \colhead{(7)} & \colhead{(8)}
}
\startdata
PG 0007+106	& $	-270	\pm	18	$ & $	668	\pm	43	$ & $	-3.26	\pm	0.18	$ &	3.28	&	4.32	&	0.033	&	--0.38	\\
PG 0050+124	& $	-32	\pm	22	$ & $	894	\pm	58	$ & $	-4.63	\pm	0.23	$ &	3.41	&	112.44	&	0.722	&	--0.76	\\
PG 0844+249	& $	158	\pm	22	$ & $	700	\pm	51	$ & $	-4.52	\pm	0.29	$ &	2.42	&	\nodata	&	\nodata	&	--0.58	\\
PG 1119+120	& $	-82	\pm	2	$ & $	25	\pm	4	$ & $	-8.49	\pm	1.25	$ &	4.73	&	\nodata	&	\nodata	&	--0.04	\\
	& $	17	\pm	27	$ & $	244	\pm	56	$ & $	-2.00	\pm	0.40	$ &	1.11	&	29.66	&	0.693	&	--0.09	\\
PG 1211+143	& $	-270	\pm	4	$ & $	466	\pm	14	$ & $	-37.80	\pm	1.88	$ &	20.43	&	5.82	&	0.065	&	--3.11	\\
	& $	-61	\pm	12	$ & $	870	\pm	12	$ & $	-36.10	\pm	1.77	$ &	19.52	&	10.38	&	0.062	&	--5.56	\\
PG 1351+640	& $	-68	\pm	18	$ & $	952	\pm	45	$ & $	-3.77	\pm	0.15	$ &	3.13	&	21.32	&	0.118	&	--0.63	\\
PG 1411+442	& $	-288	\pm	11	$ & $	735	\pm	27	$ & $	-6.19	\pm	0.20	$ &	5.21	&	\nodata	&	\nodata	&	--0.79	\\
PG 1448+273	& $	-106	\pm	29	$ & $	723	\pm	69	$ & $	-1.98	\pm	0.16	$ &	1.54	&	128.80	&	1.079	&	--0.26	\\
PG 1501+106	& $	-140	\pm	27	$ & $	550	\pm	64	$ & $	-2.15	\pm	0.22	$ &	1.47	&	$ >$155.41	&	1.966	&	--0.23	\\
F15250+3608	& $	-71	\pm	21	$ & $	858	\pm	52	$ & $	-5.76	\pm	0.29	$ &	2.73	&	60.56	&	0.386	&	--0.92	\\
PG 1617+175 (U)	& $	-95	\pm	30	$ & $	870	\pm	71	$ & $	-4.60	\pm	0.32	$ &	2.08	&	\nodata	&	\nodata	&	--0.67	\\
PG 2130+099	& $	187	\pm	35	$ & $	325	\pm	87	$ & $	-2.52	\pm	0.55	$ &	0.83	&	26.35	&	0.446	&	--0.15	\\
	& $	-247	\pm	14	$ & $	172	\pm	33	$ & $	-4.68	\pm	0.74	$ &	1.55	&	31.27	&	1.104	&	--0.15	\\
\enddata
\tablecomments{
Col.(1): Galaxy name.  Col.(2):  Heliocentric velocity measured from HI relative to the systematic velocity presented in Table~\ref{tab:sample}, as modeled using a Gaussian fit, and its associated measurement error.  Col.(3):  Full width of the absorption feature at the 50\% flux level as measured using a Gaussian model and its error.  Col.(4):  Peak absorption flux and its error as measured using a Gaussian model.  Col.(5):  Signal-to-noise ratio measured as a ratio between the peak flux and channel rms listed in Table~\ref{tab:sample}.  Col.(6):  Integrated HI column density based on the absorption model, assuming a spin temperature of 1000~K, and using Equation 4 based on the optical depth profile derived using Equation 3.  For PG~1501+106, the 1.4~GHz core flux density is only an upper limit, so its derived $N_{\rm HI}$ value is a lower limit.  Col.(7):  Peak optical depth based on Equation~3.  Col.(8):  Velocity-integrated HI flux in absorption measured from the best-fit Gaussian component models. 
}
\label{tab:complex_absorption}
\end{deluxetable}

\begin{deluxetable}{lrrrcccrrr}
\tablecolumns{10}
\tablecaption{Measurements of HI Outflow}
\tablewidth{0pt}
\tablehead{\colhead{Source} & \colhead{$V_\odot$} &\colhead{$V_{\rm \odot,~HI}$} &\colhead{$N_{\rm HI}$} & \colhead{FWZI} & \colhead{$V_{\rm max}$}&\colhead{$V_{98\%}$}& \colhead{$\dot{M}$ (wide)}&\colhead{$\dot{M}$ (jet)} &\colhead{$\log (\frac{\dot{E} {\rm (jet)}}{{\rm erg~s^{-1}}})$}\\
\colhead{} & \colhead{(km~s$^{-1}$)} &\colhead{(km~s$^{-1}$)}& \colhead{(10$^{21}$~cm~$^{-2}$)} & \colhead{(km~s$^{-1}$)} &\colhead{(km~s$^{-1}$)}&\colhead{(km~s$^{-1}$)}& \colhead{($M_\odot$~yr$^{-1}$)}&\colhead{($M_\odot$~yr$^{-1}$)} &\colhead{}\\
\colhead{(1)} & \colhead{(2)} & \colhead{(3)} & \colhead{(4)} & \colhead{(5)} &\colhead{(6)} &\colhead{(7)}&\colhead{(8)}&\colhead{(9)} &\colhead{(10)}
}
\startdata
\multicolumn{10}{c}{ULIRGs}\\ \hline	
F04103-2838 (M)	&	--409	& $	-538	\pm	36	$ &	81	& $	1267	\pm	212	$ & $	-1171	\pm	214	$ & $	-920	\pm	70	$ & 	1089	&	4.4	&	41.6	\\
F05189-2524	&	--374	& $	-374	\pm	9	$ &	10	& $	336	\pm	67	$ & $	-542	\pm	68	$ & $	-475	\pm	21	$ & 	97	&	0.4	&	40.3	\\
	&	--79	& $	-79	\pm	13	$ &	12	& $	459	\pm	90	$ & $	-309	\pm	91	$ & $	-218	\pm	29	$ & 	24	&	0.1	&	38.3	\\
\multicolumn{8}{r}{}						121$^\dag$	&	0.5$^\dag$	&		\\
Mrk 231	&	--280	& $	-280	\pm	9	$ &	2	& $	297	\pm	73	$ & $	-428	\pm	74	$ & $	-361	\pm	20	$ & 	13	&	0.1	&	39.1	\\
F15130-1958 (M)	&	--847	& $	-847	\pm	10	$ &	39	& $	266	\pm	76	$ & $	-980	\pm	77	$ & $	-923	\pm	22	$ & 	818	&	3.3	&	41.9	\\
	&	--451	& $	-451	\pm	15	$ &	61	& $	494	\pm	114	$ & $	-697	\pm	115	$ & $	-594	\pm	34	$ & 	687	&	2.7	&	41.3	\\
\multicolumn{8}{r}{}						1505$^\dag$	&	6.0$^\dag$	&		\\
F15250+3608	&	--71	& $	18	\pm	23	$ &	606	& $	2744	\pm	172	$ & $	-1354	\pm	173	$ & $	-53	\pm	28	$ & 	\nodata	&	\nodata	&	\nodata	\\
\multicolumn{3}{c}{Median}							 &	39	& $	459	\pm	90	$ & $	-697	\pm	91	$ & $	-475	\pm	28	$ & 	605	&	1.6	&	40.8	\\
\cutinhead{FIR-strong Quasars}\\
PG 0050+124	&	--32	& $	-216	\pm	30	$ &	1124	& $	2815	\pm	191	$ & $	-1623	\pm	192	$ & $	-977	\pm	58	$ & 	6072	&	24.3	&	41.6	\\
PG 1119+120	&	--82	& $	-67	\pm	4	$ &	\nodata	& $	85	\pm	17	$ & $	-110	\pm	17	$ & $	-89	\pm	5	$ & 	\nodata	&	\nodata	&	\nodata	\\
	&	17	& $	32	\pm	27	$ &	297	& $	721	\pm	182	$ & $	-329	\pm	184	$ & $	-176	\pm	55	$ & 	\nodata	&	\nodata	&	\nodata	\\
PG 1211+143	&	--270	& $	-270	\pm	17	$ &	58	& $	1680	\pm	54	$ & $	-1109	\pm	54	$ & $	-666	\pm	21	$ & 	392	&	1.6	&	40.6	\\
	&	--61	& $	-61	\pm	21	$ &	104	& $	3118	\pm	55	$ & $	-1620	\pm	56	$ & $	-801	\pm	23	$ & 	158	&	0.6	&	38.9	\\
\multicolumn{8}{r}{}						550$^\dag$	&	2.2$^\dag$	&		\\
PG 1244+026	&	--381	& $	-381	\pm	18	$ &	152	& $	696	\pm	62	$ & $	-729	\pm	63	$ & $	-573	\pm	25	$ & 	1450	&	5.8	&	41.4	\\
PG 1351+640	&	--68	& $	264	\pm	34	$ &	213	& $	2946	\pm	147	$ & $	-1210	\pm	148	$ & $	-547	\pm	51	$ & 	\nodata	&	\nodata	&	\nodata	\\
PG 1440+356	&	325	& $	325	\pm	4	$ &	\nodata	& $	1104	\pm	45	$ & $	-227	\pm	45	$ & $	41	\pm	8	$ & 	\nodata	&	\nodata	&	\nodata	\\
PG 1501+106	&	--140	& $	-201	\pm	27	$ &	$>$ 1554	& $	1637	\pm	199	$ & $	-1020	\pm	201	$ & $	-669	\pm	61	$ & 	$>$ 7813	&	$>$ 31.3	&	$>$ 41.6	\\
PG 2130+099	&	--247	& $	-314	\pm	24	$ &	313	& $	547	\pm	111	$ & $	-587	\pm	112	$ & $	-460	\pm	37	$ & 	2453	&	9.8	&	41.5	\\
\multicolumn{3}{c}{Median}							&	213	& $	1370	\pm	87	$ & $	-1020	\pm	88	$ & $	-560	\pm	31	$ & 	1450	&	5.8	&	41.5	\\
\cutinhead{FIR-weak Quasars}\\
PG 0007+106	&	--270	& $	-59	\pm	21	$ &	43	& $	2053	\pm	133	$ & $	-1085	\pm	134	$ & $	-627	\pm	42	$ & 	64	&	0.3	&	38.5	\\
PG 0844+249	&	158	& $	38	\pm	26	$ &	\nodata	& $	2198	\pm	174	$ & $	-1061	\pm	175	$ & $	-558	\pm	51	$ & 	\nodata	&	\nodata	&	\nodata	\\
PG 1411+442	&	--288	& $	-83	\pm	18	$ &	\nodata	& $	2362	\pm	91	$ & $	-1265	\pm	92	$ & $	-709	\pm	29	$ & 	\nodata	&	\nodata	&	\nodata	\\
PG 1448+273	&	--106	& $	-38	\pm	31	$ &	1288	& $	2129	\pm	215	$ & $	-1102	\pm	217	$ & $	-653	\pm	67	$ & 	\nodata	&	\nodata	&	\nodata	\\
PG 1617+175 (U)	&	--95	& $	-7	\pm	33	$ &	\nodata	& $	2737	\pm	234	$ & $	-1376	\pm	236	$ & $	-748	\pm	69	$ & 	\nodata	&	\nodata	&	\nodata	\\
\multicolumn{3}{c}{Median}							&	666	& $	2198	\pm	174	$ & $	-1102	\pm	175	$ & $	-653	\pm	51	$ & 	64	&	0.3	&	38.5	\\

\enddata
\tablecomments{
$^\dag$  The sum of the mass outflow rates from the multiple Gaussian components in the galaxy.  Col.(1): Galaxy name.  Col.(2):  Heliocentric velocity of the absorption component as tabulated in Tables~\ref{tab:absorption} and \ref{tab:complex_absorption}. Col. (3):  With the exception of PG~1211+143 and PG~1440+356, this is the outflow velocity measured relative to the HI systemic velocity measured from HI emission using values tabulated in column (2) of Table~\ref{tab:emission} or \ref{tab:complex_emission}.  $V_{\rm \odot, ~HI} = V_\odot$ if no emission is detected.  For PG~1211+143 and PG~1440+356, this velocity is relative to the systemic velocity measured from optical emission lines, since the detection of HI emission is uncertain (Figures~\ref{fig:pg1211var} and \ref{fig:pg1440var}).  Col. (4): Integrated HI column density based on the absorption model from Tables~\ref{tab:absorption} and \ref{tab:complex_absorption}.  Col. (5):  Full-width at zero intensity based on the best-fit Gaussian absorption model and its associated error.   Col. (6):  The maximum outflow velocity where $V_{\rm max} \equiv V_{\rm \odot,~HI} - \frac{1}{2}$FWZI.  Col. (7):  The velocity at which 98\% of the gas has lower velocity where $V_{98\%} \equiv V_{\rm \odot,~HI} - 2 (\frac{\rm FWHM}{2.35})$.  Col. (8):  Derived mass outflow rate calculated using Equation~5 and assuming the outflow is in a wide-angled cone where the mass is flowing into a solid angle of $\pi$ from a radius of 1~kpc following the same assumptions as \citet{morganti05b}.  Col. (9):  Same as column~(8) but for a jet whose opening angle is assumed to be 5$^\circ$ launched at a distance of 0.5~kpc.  Col. (10):  Log of the kinetic energy outflow rate as calculated using Equation~6 for the jet scenario.  
}
\label{tab:outflows}
\end{deluxetable}

\newpage


\end{document}